\documentclass[12pt,a4paper]{iopart}

\usepackage{bbm}
\usepackage{graphicx}
\usepackage{oldgerm}
\usepackage{mathpak}
\usepackage{amssymb}
\usepackage{amsthm}
\usepackage{lscape}

\newtheorem{thm}{Theorem}
\newcommand{\bra}[1]{\ensuremath{\left\langle #1\right|}}
\newcommand{\ket}[1]{\ensuremath{\left|#1\right\rangle}}

\newcommand{\spann}{\textrm{span}}

\begin{document}
\begin{flushright}
UWThPh-2011-08
\end{flushright}

\title{Composite parameterization and Haar measure for all unitary and special unitary groups}

\author{Christoph Spengler, Marcus Huber, Beatrix C. Hiesmayr}
\address{Faculty of Physics, University of Vienna, Boltzmanngasse 5, 1090 Vienna, Austria}
\ead{Christoph.Spengler@univie.ac.at}

\begin{abstract}
We adopt the concept of the composite parameterization of the unitary group $\mathcal{U}(d)$ to the special unitary group $\mathcal{SU}(d)$. Furthermore, we also consider the Haar measure in terms of the introduced parameters. We show that the well-defined structure of the parameterization leads to a concise formula for the normalized Haar measure on $\mathcal{U}(d)$ and $\mathcal{SU}(d)$. With regard to possible applications of our results, we consider the computation of high-order integrals over unitary groups.
\end{abstract}


\maketitle

\section{Introduction}
Unitary and special unitary groups play an important role in various fields of physics. Several problems that arise in the context of these groups require to express them in terms of a set of real parameters. In general, such parameterizations are not unique in the sense that the considered groups can be parameterized in many different ways. With regard to the diversity of problems it is reasonable to have a repertoire of different parameterizations available, in order to be able to choose the most convenient one for a given problem. In Ref.~\cite{Composite} we recently introduced the composite parameterization of the unitary group $\mathcal{U}(d)$ as an alternative to the canonical parameterization $U=\exp(i H)$ via hermitian matrices $H$ and those presented in Refs.~\cite{TilmaUN,Jarlskog,Dita}. It was shown that our parameterization enables a simple identification of redundant parameters when it is applied to describing orthonormal bases, density matrices of arbitrary rank and subspaces. For all these objects we found representations containing the minimal number of parameters needed. Due to its concise notation, simple implementation and computational benefits it has already found widespread applications in research on lattice correlation functions \cite{RWJohnson1}, quantum nonlocality \cite{geomNL}, genuine multipartite entanglement \cite{GenMultClass,Ma,RelClasses} and quantum secret sharing \cite{Schauer}.

The aim of the present paper is twofold. First, we adopt our concept that was used in Ref.~\cite{Composite} to obtain a novel parameterization of the special unitary group $\mathcal{SU}(d)$. The need of additional representations of this group is not only apparent because of its vital relevance in all kinds of fields involving quantum physics (see Ref.~\cite{Redkov} for an overview), but moreover is given because the number of available parameterizations is relatively low compared to $\mathcal{U}(d)$. Here, our parameterization is proposed as an alternative to the canonical parameterization $U=\exp(i H)$ via \emph{traceless} hermitian matrices $H$ and the generalized Euler angle parameterization \cite{TilmaSUN} introduced by Tilma and Sudarshan.

Second, we rigorously derive the normalized Haar measure in terms of the introduced parameters for both the unitary $\mathcal{U}(d)$ and the special unitary group $\mathcal{SU}(d)$ of arbitrary dimension. In form of an infinitesimal volume element of a group, the Haar measure contains all information about the distribution density in its parameter representation. This in return is essential for the capability of generating uniformly distributed random unitaries, density matrices and subspaces, as they are important in, for instance, Monte Carlo simulations \cite{Fresch} or quantum data hiding \cite{Terhal,Winter1}. Explicit expressions of the normalized Haar measure can furthermore be useful to tackle group integrals as they appear in lattice QCD \cite{Verbaaschot}, quantum optics \cite{Serafini}, stochastic processes \cite{Katori} and mesoscopic systems \cite{Savin}. In quantum information they can be found in the context of symmetric states (\emph{Werner} and isotropic states \cite{Werner,Eggeling,Chruscinski}) and the \emph{a priori entanglement} of quantum systems \cite{SlaterA,HaydenEnt,Gross}.

Within recent years much attention has been paid to integrals over unitary groups, i.e. $\int_{\mathcal{U}(d)} f(U,U^*) dU $ whose integrand is a polynomial in $U$ and $U^{*}$. It was shown that such integrals can be replaced by a sum of function values $f(U_i,U_i^{*})$ using a finite set of unitaries $\{ U_i \}_N$. Such sets are termed \emph{unitary $t$-designs} \cite{Gross,Dankert,Winter2}, wherein $t$ denotes that the degree of the polynomial in $U$ and $U^{*}$ is at most $t$. Whereas the existence of a unitary $t$-design was proven for all dimensions $d$ and all polynomial degrees $t$ (see Ref.~\cite{Seymour}), it is generally unknown how to construct them for arbitrary $d$ and $t$. In addition, there currently exists no simple analytic method for solving any integral over any polynomial. Due to this, there have been several attempts to find schemes to approximate unitary $t$-designs. Here, the usefulness of our results in the context of unitary $t$-designs and their approximations is obvious: Since we do not only provide the Haar measure but also the exact parameter ranges for group covering, the integration of polynomials can be performed explicitly and solved analytically in many cases. This in return allows to verify (or falsify) suggested unitary designs and to test the accuracy of approximations. Apart from that, our tools for computing integrals are not limited to polynomials, but are applicable to arbitrary functions.

The paper is organized as follows. In Section \ref{compU} we review the composite parameterization of the unitary group $\mathcal{U}(d)$. In Section \ref{compSU} we introduce the composite parameterization of the special unitary group $\mathcal{SU}(d)$. In Section \ref{haarU} the general formula for the Haar measure on $\mathcal{U}(d)$ is stated and proven. An analogous formula for the Haar measure on the special unitary group $\mathcal{SU}(d)$ can be found in Section \ref{haarSU}. Finally, in Section \ref{NotesInt} we make useful remarks on computing integrals over $\mathcal{U}(d)$ and $\mathcal{SU}(d)$ using the composite parameterization and the associated Haar measure.
\section{Composite parameterization of the unitary group $\mathcal{U}(d)$}
\label{compU}
Consider a $d$-dimensional $(d\geq2)$ complex Hilbert space $\mathcal{H}=\mathbb{C}^d$ spanned by the orthonormal basis $\{ \ket{1},\ldots,\ket{d} \}$. On this space define $d$ one-dimensional projectors
\begin{align}
\label{projector}
P_l=\ket{l}\bra{l} \hspace{4.7cm} 1 \leq l \leq d
\end{align}
and ${d(d-1)}/{2}$ anti-symmetric matrices \footnote{These can be considered as generalizations of the Pauli matrix $\sigma_y$}
\begin{align}
\label{antisymm}
Y_{m,n}=-i\ket{m} \bra{n} + i \ket{n} \bra{m} \hspace{1.2cm} 1\leq m < n \leq d
\end{align}
each acting on a two-dimensional subspace spanned by $\ket{m}$ and $\ket{n}$. In our previous paper \cite{Composite}, using these operators we have proven the following theorem:
\begin{samepage}
\begin{thm}
\label{theorem1}
Any operator of the unitary group $\mathcal{U}(d)$ can be written as \footnote{The order of the product is $\prod_{i=1}^{N}A_i=A_1 \cdot A_{2} \cdots A_N$}
\begin{align}
\label{Uc}
U_C=\left[\prod_{m=1}^{d-1} \left(\prod_{n=m+1}^{d} \exp \left( i P_n \lambda_{n,m} \right) \exp \left( i Y_{m,n} \lambda_{m,n} \right)  \right) \right] \cdot \left[ \prod_{l=1}^{d} \exp(i P_l \lambda_{l,l})\right] \ ,
\end{align}
using $d^2$ real parameters $\{ \lambda_{m,n} \}_{m,n=1,\ldots,d}$ in the ranges $\lambda_{m,n} \in \left[0, 2 \pi \right]$ for $m \geq n$ and $\lambda_{m,n} \in \left[0, \frac{\pi}{2} \right]$ for $m < n$.
\end{thm}
\end{samepage}
The idea behind this construction was to compose the unitary group out of `elementary operations' such as rotations and phase shifts. Here, these operations are realized through the matrix exponential functions $\exp \left( i Y_{m,n} \alpha \right)$ (generates a rotation) and $\exp \left( i P_l \alpha \right)$ (shifts a phase). To make an ansatz for an arbitrary unitary operator it is useful to think of unitaries as orthonormal basis transformations. In this way, the form (\ref{Uc}) can be interpreted as one option of incorporating $d$ global phase operations
\begin{align}
\exp(i P_l \lambda_{l,l})
\end{align}
as well as $d(d-1)/2$ rotations followed by relative phase shifts
\begin{align}
\label{distinc}
\Lambda_{m,n}=\exp \left( i P_n \lambda_{n,m} \right) \exp \left( i Y_{m,n} \lambda_{m,n} \right)
\end{align}
that result from partitioning the Hilbert space into two-dimensional subspaces according to (\ref{antisymm}). That this $d^2$ parameter set of unitary operators indeed covers the whole unitary group $\mathcal{U}(d)$ was proven in Ref.~\cite{Composite}. This was done by showing that for any $U\in\mathcal{U}(d)$ there exists a $U_C$ such that $U_C^{\dagger} U=\mathbbm{1}$.

\section{Composite parameterization of the special unitary group $\mathcal{SU}(d)$}
\label{compSU}
Unitary operators obeying $\det U =1$ constitute a subgroup of $\mathcal{U}(d)$ called the special unitary group $\mathcal{SU}(d)$. The first new result of this paper is that this group can be parameterized similarly as $\mathcal{U}(d)$ using the concept of the composite parameterization. First, let us note that as special unitary operators satisfy an additional constraint the special unitary group $\mathcal{SU}(d)$ can be described by $d^2-1$ real parameters. As redundant parameters are undesirable we have to find a parameterization that contains exactly this number of parameters. Second, it is known that $U=\exp(i H \alpha)$ is special unitary for all $\alpha \in \mathbb{R}$ only if $H$ is hermitian and \emph{traceless}. In the composite parameterizaton of the unitary group $\mathcal{U}(d)$ we have used matrix exponentials of $P_l$ and $Y_{m,n}$ as defined in Eq.~(\ref{projector}) and Eq.~(\ref{antisymm}). The latter is already traceless but any one-dimensional projector $P_l$ has $\Tr(P_l)=1$. As the projectors $P_l$ were used to create phase shifts it is clear that they have to be replaced by a set of diagonal \emph{traceless} operators. For that we introduce the following operators
\begin{align}
\label{genZ}
Z_{m,n}=\ket{m} \bra{m} - \ket{n} \bra{n} \hspace{1.2cm} 1\leq m < n \leq d \ .
\end{align}
These are possible generalizations of the diagonal Pauli matrix $\sigma_z$ acting on the subspace spanned by $\ket{m}$ and $\ket{n}$. In (\ref{distinc}) the operation $\exp \left( i P_n \lambda_{n,m} \right)$ was used to generate a relative phase shift between the vector components $\ket{m}$ and $\ket{n}$. However, the same effect can also be achieved with $\exp \left( i Z_{m,n} \lambda_{n,m} \right)$ meaning that (\ref{distinc}) can be replaced by
\begin{align}
\label{distincSU}
\Lambda_{m,n}=\exp \left( i Z_{m,n} \lambda_{n,m} \right) \exp \left( i Y_{m,n} \lambda_{m,n} \right) \ .
\end{align}
It now remains to turn our attention to the last part of (\ref{Uc}), i.e. $\left[ \prod_{l=1}^{d} \exp(i P_l \lambda_{l,l})\right]$. Here, each $\exp(i P_l \lambda_{l,l})$ is regarded as a global phase operation on $\ket{l}$. Special unitarity implies that there are only $d-1$ independent global phase operations instead of $d$ for $\mathcal{U}(d)$. There is no unique way how these operations can be realized using matrix exponentials of (\ref{genZ}). However, a possible and convenient choice is
\begin{align}
\exp(i Z_{l,d} \lambda_{l,l}) \hspace{1.8cm} 1 \leq l \leq d-1 \ .
\end{align}
In this version, the first $d-1$ vectors $\ket{1},\ldots,\ket{d-1}$ experience the phase shifts\\$e^{i \lambda_{1,1} }\ket{1},\ldots,e^{i \lambda_{d-1,d-1} }\ket{d-1}$ while the last vector $\ket{d}$ gets phase shifted in the overall inverse direction, i.e. $e^{-i \sum_{l=1}^{d-1}\lambda_{l,l}}\ket{d}$. Note that according to our labeling there is no parameter $\lambda_{d,d}$. Thus, in total we have the desired number of $d^2-1$ parameters $\lambda_{m,n}$. In summary, this leads to the next theorem:
\begin{samepage}
\begin{thm}
\label{theorem2}
Any operator of the special unitary group $\mathcal{SU}(d)$ can be written as
\begin{align}
\label{SUc}
U_{C}=\left[\prod_{m=1}^{d-1} \left(\prod_{n=m+1}^{d} \exp \left( i Z_{m,n} \lambda_{n,m} \right) \exp \left( i Y_{m,n} \lambda_{m,n} \right)  \right) \right] \cdot \left[ \prod_{l=1}^{d-1} \exp(i Z_{l,d} \lambda_{l,l})\right] \ ,
\end{align}
using $d^2-1$ real parameters $\{ \lambda_{m,n} \}$\footnote{Note that the indices $m$ and $n$ again run from $1$ to $d$ except that there is no $\lambda_{d,d}$.} in the ranges $\lambda_{m,n} \in \left[0, \pi \right]$ for $m > n$, $\lambda_{m,n} \in \left[0, \frac{\pi}{2} \right]$ for $m < n$ and $\lambda_{m,n} \in \left[0, 2\pi \right]$ for $m = n$.
\end{thm}
\end{samepage}

\begin{proof}
Proving that any $U\in \mathcal{SU}(d)$ may be written as (\ref{SUc}) is equivalent to showing that $U_C^{\dagger}U=\mathbbm{1}$ can be achieved for all group elements. Let $U=\sum_{r,s=1}^{d} a_{r,s} \ket{r}\bra{s}$ be an arbitrary special unitary operator, i.e. $U$ fulfils $\sum_{i=1}^{d} a_{m,i}^{*}a_{n,i}=\sum_{i=1}^{d} a_{i,m}^{*}a_{i,n}=\delta_{mn}$ and $\det U=1$.
The conjugate transpose of $U_C$ as given in (\ref{SUc}) using the abbreviation (\ref{distincSU}) is
\begin{align}
U_C^{\dagger}=\left[ \prod_{l=1}^{d-1} \mbox{exp}(-i Z_{d-l,d} \lambda_{d-l,d-l})\right] \cdot \left[\prod_{m=1}^{d-1} \left(\prod_{n=1}^{m}  \Lambda_{d-m,d+1-n}^{\dagger}  \right) \right] \ .
\end{align}
The order of the factors in $U_C^{\dagger}$ implies that $\Lambda_{1,2}^{\dagger}$ acts first on $U$. For $U'=\Lambda_{1,2}^{\dagger} U=\sum_{r,s=1}^{d} a'_{r,s} \ket{r}\bra{s}$ one obtains
\begin{align}
\label{proofa}
a'_{1,s}&=e^{-i\lambda_{2,1}}\cos(\lambda_{1,2})a_{1,s}-e^{i\lambda_{2,1}}\sin(\lambda_{1,2})a_{2,s} \ , \\
a'_{2,s}&=e^{-i\lambda_{2,1}}\sin(\lambda_{1,2})a_{1,s}+e^{i\lambda_{2,1}}\cos(\lambda_{1,2})a_{2,s} \ .
\end{align}
All other components remain unchanged, i.e. $a'_{r,s}=a_{r,s}$ for $r>2$ if $d>2$. We observe that $a'_{2,1}$ can always be made zero using particular values for $\lambda_{1,2}$ and $\lambda_{2,1}$: In case $a_{1,1}$ and $a_{2,1}$ both are zero, both parameters $\lambda_{1,2}$ and $\lambda_{2,1}$ can be chosen freely. If only $a_{1,1}=0$ one chooses $\lambda_{1,2}=\frac{\pi}{2}$. If both are unequal zero then $a'_{2,1}$ vanishes for
\begin{align}
\arg(e^{i2\lambda_{2,1}}a_{2,1})&=\arg(-a_{1,1}) \ ,\\
\tan(\lambda_{1,2})&=\frac{|a_{2,1}|}{|a_{1,1}|} \ .
\label{proofb}
\end{align}
This can always be achieved with $\lambda_{2,1}\in [0,\pi]$ and $\lambda_{1,2}\in [0,\frac{\pi}{2}]$. Analogously, one can make the component $a''_{3,1}$ of $U''=\Lambda^{\dagger}_{1,3}U'$ zero for the case $d>2$. In this way, all components $\overline{a}_{r,s}$ with $r>s$ of $\overline{U}=\prod_{m=1}^{d-1} \left(\prod_{n=1}^{m}  \Lambda_{d-m,d+1-n}^{\dagger}  \right) U=\sum_{r,s=1}^{d} \overline{a}_{r,s} \ket{r}\bra{s}$ can be made zero. The unitarity constraints $\sum_{i=1}^{d} \overline{a}_{m,i}^{*}\overline{a}_{n,i}=\sum_{i=1}^{d} \overline{a}_{i,m}^{*}\overline{a}_{i,n}=\delta_{mn}$ imply that also all $\overline{a}_{r,s}$ with $r<s$ have become zero during this procedure. Hence, $\overline{U}$ is diagonal $\overline{U}=\sum_{r=1}^{d} \overline{a}_{r,r} \ket{r}\bra{r}$ where $\overline{a}_{r,r}$ are complex numbers of magnitude one, i.e. $\overline{a}_{r,r}=e^{i\alpha_r}$. The first $d-1$ phases $\alpha_1,\ldots,\alpha_{d-1}$ can be compensated via $\left[ \prod_{l=1}^{d-1} \mbox{exp}(-i Z_{d-l,d} \lambda_{d-l,d-l})\right] \overline{U}$ by choosing $\lambda_{r,r}=\alpha_r$. This is guaranteed to be achievable with $\lambda_{r,r}\in [0, 2\pi ]$. Recall that so far we have only multiplied special unitary operators implying that $U_C^{\dagger}U=\left[ \prod_{l=1}^{d-1} \mbox{exp}(-i Z_{d-l,d} \lambda_{d-l,d-l})\right] \overline{U}$ is still a member of $\mathcal{SU}(d)$. We have achieved that $U_C^{\dagger}U$ is diagonal and that the first $d-1$ diagonal entries are all equal $1$. Now, the fact that $\det(U_C^{\dagger}U) =1$ still holds implies that also the last diagonal entry equals $1$. Hence, $U_C^{\dagger}U=\mathbbm{1}$, which proves the theorem.
\end{proof}

\subsection{Remarks on the composite parameterzation}
\label{remarksUSU}
It is worth mentioning some properties of the composite parameterization. In our previous paper \cite{Composite} on the parameterization of $\mathcal{U}(d)$ we have shown that it can be rather insightful to gather the parameters $\lambda_{m,n}$ in a matrix
\begin{align}
\begin{array}{c}
\ \\
\ \\
\mbox{relative phases} \rightarrow
\end{array}
 \left[
  \begin{array}{ccc}
    \lambda_{1,1} & \cdots & \lambda_{1,d} \\
    \vdots & \ddots & \vdots \\
    \lambda_{d,1} & \cdots & \lambda_{d,d} \\
  \end{array}
\right]
\begin{array}{c}
\leftarrow \mbox{rotations} \\
\ \\
\
\end{array} \ .
\end{align}
In this notation the lower left entries represent relative phase shifts, the diagonal global phase shifts and the upper right rotations (with respect to the computational basis $\{ \ket{1},\ldots,\ket{d} \}$). Using this representation is particularly useful for illustrating which parameters are irrelevant for certain tasks. For instance, we have shown that for parameterizing an orthonormal set of $k$ vectors $\{ \ket{\Psi_1}, \ldots, \ket{\Psi_k} \}$ only the $k(2d-k-1)$ non-zero parameters of the following matrix are relevant
\begin{align}
& \left[
  \begin{array}{ccccccc}
    0 & \lambda_{1,2} & \cdots & \lambda_{1,k+1} & \cdots & \lambda_{1,d} \\
    \lambda_{2,1} & \ddots & \ddots & \vdots & \ddots & \vdots \\
    \vdots & \ddots& 0 &  \lambda_{k,k+1} & \cdots & \lambda_{k,d} \\
         \lambda_{k+1,1} & \cdots & \lambda_{k+1,k}   & 0 & \cdots & 0 \\
     \vdots & \ddots & \vdots   & \vdots & \ddots & \vdots\\
     \lambda_{d,1} & \cdots & \lambda_{d,k}   & 0 & \cdots & 0 \\
   \end{array}
\right]
\begin{array}{c}
  \left. {\begin{array}{c}
  \\
  \\
  \\
    \end{array}} \right\} k \hspace{7mm} \\
    \\
    \left. {\begin{array}{c}
  \\
  \\
  \\
  \end{array}} \right\} d-k
\end{array} \\
& \hspace{0.6cm} \underbrace{ \hspace{3.1cm} }_{k} \hspace{7mm} \underbrace{ \hspace{2.8cm} }_{d-k} \hspace{2cm} . \nonumber
\end{align}
A similar example was given for parameterizing the set of $k$-dimensional subspaces in $\mathbb{C}^d$ where the corresponding $2k(d-k)$ relevant parameters $\lambda_{m,n}$ were illustrated. For the composite parameterization of the special unitary group $\mathcal{SU}(d)$ all this remains valid as the operations (\ref{distinc}) and (\ref{distincSU}) are equivalent up to a global phase and are applied in the same order in both cases. Using this matrix representation for the special unitary group one should only keep in mind that there is no diagonal element $\lambda_{d,d}$.

Let us illustrate another important feature of the composite parameterization. The $(d-k)^2$ or $(d-k)^2-1$, respectively, non-zero parameters
\begin{align}
& \left[
  \begin{array}{ccccccc}
    0 & 0 & \cdots & 0 & \cdots & 0 \\
    0 & \ddots & \ddots & \vdots & \ddots & \vdots \\
    \vdots & \ddots& 0 &  0 & \cdots & 0 \\
         0 & \cdots & 0   & \lambda_{k+1,k+1} & \cdots & \lambda_{k+1,d}  \\
     \vdots & \ddots & \vdots   & \vdots & \ddots & \vdots\\
     0 & \cdots & 0   & \lambda_{d,k+1} & \cdots & \lambda_{d,d} \\
   \end{array}
\right]
\begin{array}{c}
  \left. {\begin{array}{c}
  \\
  \\
  \\
    \end{array}} \right\} k \hspace{7mm} \\
    \\
    \left. {\begin{array}{c}
  \\
  \\
  \\
  \end{array}} \right\} d-k
\end{array} \\
& \hspace{0.6cm} \underbrace{ \hspace{1.9cm} }_{k} \hspace{8mm} \underbrace{ \hspace{3.1cm} }_{d-k} \hspace{2cm}  \nonumber
\end{align}
correspond to the (special) unitary group for the $(d-k)$-dimensional subspace defined by $\spann(\ket{k+1},\ldots,\ket{d})$. This directly follows from the correct number of required parameters in combination with the fact that the subspace $\spann(U_C\ket{1},\ldots,U_C\ket{k})$ is independent of the illustrated parameters. Note that this feature will be helpful in an upcoming proof.
\section{Haar measure on the unitary group $\mathcal{U}(d)$}
\label{haarU}
Let us now assign an infinitesimal volume element $dU_d$ to the unitary group $\mathcal{U}(d)$ in terms of the composite parameterization $U_C=U_C(\lambda_{1,1},\ldots,\lambda_{d,d})$. This can be achieved by determining the associated Haar measure. This means that, as for any compact Lie group, we must find a measure of volume which is left and right invariant \cite{Haarorg,Folland}. Explicitly, we require that $dU_d$ satisfies
\begin{align}
dU_d=d(U_C) = d(U_1 U_C) = d(U_C U_2) \ ,
\end{align}
for all $U_1, U_2 \in \mathcal{U}(d)$. Generally, an invariant measure is (up to an irrelevant constant) determined by the absolute value of the Jacobian determinant
\begin{align}
J_d= \left| \det (j_{k,l}) \right|=\left| \det \frac{\partial(u_1,\ldots,u_{d^2})}{\partial (\alpha_1,\ldots,\alpha_{d^2})} \right|  \ ,
\end{align}
wherein $\{ u_k \}$ are coefficients of the unitary $U=U(\alpha_1,\ldots,\alpha_{d^2})$ expanded in an orthogonal operator basis $\{ b_k \}$ of $\mathbb{C}^d \times \mathbb{C}^d$, i.e.
\begin{align}
u_k=\frac{\Tr(b_k^{\dagger}U)}{\Tr(b_k^{\dagger}b_k)}  \hspace{1.8cm} \left(  \Rightarrow U=\sum_{k=1}^{d^2} u_k b_k \right) \ ,
\end{align}
and where $\{ \alpha_l \}$ are any $d^2$ parameters that cover $\mathcal{U}(d)$. Here, a left or right translation $U_1, U_2 \in \mathcal{U}(d)$ merely induces a unitary basis transformation of $\{ b_k \}$ which is length and angle preserving. Hence, $J_d$ is invariant under these transformations and
\begin{align}
dU_d=J_d \prod_{l=1}^{d^2} d\alpha_l
\end{align}
is a Haar measure\footnote{Intuitively: When changing from the orthogonal coordinates $\{ u_k \}$ to the non-orthogonal coordinates $\{ \alpha_l \}$ the infinitesimal volume element transforms as $\prod_{k=1}^{d^2}du_k=\left| \det \frac{\partial(u_1,\ldots,u_{d^2})}{\partial (\alpha_1,\ldots,\alpha_{d^2})} \right| \prod_{l=1}^{d^2} d\alpha_l$ according to the Jacobian determinant.}. Our aim is to derive a general expression of $dU_d$ for arbitrary $d$ in terms of the parameterization introduced in Section \ref{compU}, i.e.
\begin{align}
\label{Haargeneral}
dU_d=\frac{J_d}{N_d} \prod_{k,l=1}^{d}d\lambda_{k,l} \ ,
\end{align}
where $N_d$ is a normalization constant such that $\int_{\mathcal{U}(d)} dU_d =1$. We obtain the following:
\begin{samepage}
\begin{thm}
\label{theorem3}
In terms of the $d^2$ parameters $\lambda_{m,n}$ introduced in Theorem \ref{theorem1} the (normalized) Haar measure on the unitary group $\mathcal{U}(d)$ reads
\begin{align}
\label{HaarU}
dU_d=\frac{1}{N_d}\prod_{m=1}^{d-1}\prod_{n=m+1}^{d}\sin(\lambda_{m,n})\cos^{2(n-m)-1}(\lambda_{m,n}) \prod_{k,l=1}^{d}d\lambda_{k,l} \ ,
\end{align}
with
\begin{align}
\label{normiconstanti}
N_d=\frac{(2\pi)^{d(d+1)/2}}{\prod_{m=1}^{d-1}\prod_{n=m+1}^{d} 2(n-m)}
\end{align}
such that $\int_{\mathcal{U}(d)}dU_d=1$.
\end{thm}
\end{samepage}
\begin{proof}
For simplicity, let us start with the case $d=2$. Using the canonical operator basis $\{b_k\}=\{ \ket{1}\bra{1},\ket{1}\bra{2},\ket{2}\bra{1},\ket{2}\bra{2} \}$ and the order of the parameters $\{\alpha_l\}=\{\lambda_{1,1},\lambda_{1,2},\lambda_{2,1},\lambda_{2,2}\}$ we obtain the Jacobian matrix
\begin{align}
\label{jacobianU2}
&\frac{\partial(u_1,u_2,u_3,u_4)}{\partial (\lambda_{1,1},\lambda_{1,2},\lambda_{2,1},\lambda_{2,2})}=\\
&\left(
\begin{array}{cccc}
 i e^{i \lambda_{1,1}} \cos \lambda_{1,2} & -e^{i\lambda_{1,1}} \sin \lambda_{1,2} & 0 & 0 \\
 0 & e^{i \lambda_{2,2}} \cos \lambda_{1,2} & 0 & i e^{i \lambda_{2,2}} \sin \lambda_{1,2} \\
 -i e^{i\lambda_{1,1}+i\lambda_{2,1}} \sin \lambda_{1,2} & -e^{i\lambda_{1,1}+i\lambda_{2,1}} \cos
   \lambda_{1,2} & -i e^{i\lambda_{1,1}+i\lambda_{2,1}} \sin \lambda_{1,2} & 0 \\
 0 & -e^{i\lambda_{2,1}+i\lambda_{2,2}} \sin \lambda_{1,2} & i e^{i\lambda_{2,1}+i\lambda_{2,2}} \cos
   \lambda_{1,2} & i e^{i\lambda_{2,1}+i\lambda_{2,2}} \cos \lambda_{1,2}
\end{array}
\right) . \nonumber
\end{align}
Using the Laplace expansion and elementary simplifications one finds
\begin{align}
J_2=2\sin(\lambda_{1,2})\cos(\lambda_{1,2})=\sin(2\lambda_{1,2}) \ .
\end{align}
The relation (\ref{Haargeneral}) combined with normalization
\begin{align}
\int_{\lambda_{2,2}=0}^{2\pi}\int_{\lambda_{2,1}=0}^{2\pi}\int_{\lambda_{1,2}=0}^{\pi/2}\int_{\lambda_{1,1}=0}^{2\pi} \frac{J_2}{N_2} d\lambda_{1,1}d\lambda_{1,2}d\lambda_{2,1}d\lambda_{2,2}=1 \ ,
\end{align}
yields the normalized Haar measure
\begin{align}
\label{U2}
dU_2=\frac{1}{4\pi^3}\sin(\lambda_{1,2})\cos(\lambda_{1,2}) d\lambda_{1,1}d\lambda_{1,2}d\lambda_{2,1}d\lambda_{2,2}\ ,
\end{align}
which is in accordance with Theorem \ref{theorem3}.

One could in principle compute $dU_d$ analogously for arbitrary $d$, i.e. using the canonical operator basis $\{b_k\}=\{\ket{1}\bra{1},\ket{1}\bra{2},\ket{1}\bra{3},\ldots,\ket{d}\bra{d-1},\ket{d}\bra{d}\}$ and the naive order $\{\alpha_l\}=\{\lambda_{1,1}, \lambda_{1,2}, \lambda_{1,3},\ldots, \lambda_{d,d-1}, \lambda_{d,d}\}$. For instance, a long and cumbersome computation reveals that
\begin{align}
\label{U3}
dU_3=\frac{1}{4\pi^6}\sin(\lambda_{1,2})\cos(\lambda_{1,2})\sin(\lambda_{1,3})\cos^3(\lambda_{1,3})\sin(\lambda_{2,3})\cos(\lambda_{2,3})\prod_{k,l=1}^{3}d\lambda_{k,l}
\end{align}
demonstrating the validity of Theorem \ref{theorem3} for $d=3$. Unfortunately, in this way, computing the determinant of the $d^2 \times d^2$ Jacobian matrix becomes increasingly unfeasible the larger $d$ gets. More importantly, this approach is not suitable to prove a general expression such as (\ref{HaarU}) for all $d$. However, the Jacobian matrix can be considerably simplified by taking into account the invariance of the Haar measure, the structure of the composite parameterization as well as the freedom in the choice of the operator basis and the order of the derivatives $\partial / \partial \lambda_{x,y}$. In this way, the correctness of Theorem \ref{theorem3} can be verified for all $d$.

First, due to the left invariance of the Jacobian determinant $J_d=\left| \det (j_{k,l}) \right|$ we are allowed to perform any transformation
\begin{align}
j_{k,l}=\frac{\partial u_k}{\partial \alpha_l}=\frac{\Tr(b_k^{\dagger} \frac{\partial U}{\partial \alpha_l})}{\Tr(b_k^{\dagger}b_k)} \hspace{0.5cm} \longrightarrow \hspace{0.5cm} j'_{k,l}=\frac{\Tr(b_k^{\dagger} U_1 \frac{\partial U}{\partial \alpha_l})}{\Tr(b_k^{\dagger}b_k)} \ ,
\end{align}
with $U_1 \in \mathcal{U}(d)$. Here, it is beneficial to choose $U_1=-iU_C^{\dagger}$ since the matrix $-iU_C^{\dagger} \frac{\partial U_C}{\partial \lambda_{x,y}}$ has a simpler form due to the fact that $U_C^{\dagger}$ and $\frac{\partial U_C}{\partial \lambda_{x,y}}$ cancel each other out partially. For instance, since all projectors $P_l=\ket{l}\bra{l}$ commute, for any derivative with respect to a global phase transformation $\partial / \partial \lambda_{l,l}$ one obtains
\begin{align}
\label{diagglobal}
-iU_C^{\dagger} \frac{\partial U_C}{\partial \lambda_{l,l}}
=-iU_C^{\dagger}U_C i P_l
=P_l=\ket{l}\bra{l} \ .
\end{align}
As can directly be inferred from the structure of $U_C$ (\ref{Uc}), the derivatives $\partial U_C / \partial \lambda_{x,y}$ are
\begin{align*}
\left[\prod_{m=1}^{x-1} \prod_{n=m+1}^{d} \Lambda_{m,n}   \right] \left[\prod_{n=x+1}^{y} \Lambda_{x,n}  \right] iY_{x,y}\left[\prod_{n=y+1}^{d} \Lambda_{x,n}  \right]\left[\prod_{m=x+1}^{d-1} \prod_{n=m+1}^{d} \Lambda_{m,n}   \right] \prod_{l=1}^{d} \exp(i P_l \lambda_{l,l})
\end{align*}
for $x<y$, and
\begin{align*}
\left[\prod_{m=1}^{y-1} \prod_{n=m+1}^{d} \Lambda_{m,n}   \right] \left[\prod_{n=y+1}^{x-1} \Lambda_{y,n}  \right] iP_{x}\left[\prod_{n=x}^{d} \Lambda_{y,n}  \right]\left[\prod_{m=y+1}^{d-1} \prod_{n=m+1}^{d} \Lambda_{m,n}   \right] \prod_{l=1}^{d} \exp(i P_l \lambda_{l,l})
\end{align*}
for $x>y$. Consequently, if we apply $-iU_C^{\dagger}$ from the left, the products to the left of $iY_{x,y}$ (respectively $iP_{x}$) cancel out and we get
\begin{align}
\label{derivs}
-iU_C^{\dagger}\frac{\partial U_C}{\partial \lambda_{x,y}}\, = \,\left\{\begin{array}{ll} U_{x,y}^{\dagger}Y_{x,y}U_{x,y} \hspace{1.3cm} & \mbox{for} \ \  x<y \ , \\
		U_{x,y}^{\dagger}P_{x}U_{x,y} & \mbox{for} \ \ x>y \ , \end{array}\right.
\end{align}
where
\begin{align}
\label{drotU}
U_{x,y}\, = \,\left\{\begin{array}{ll} \left[\prod_{n=y+1}^{d} \Lambda_{x,n}  \right]\left[\prod_{m=x+1}^{d-1} \prod_{n=m+1}^{d} \Lambda_{m,n}   \right] \prod_{l=x}^{d} \exp(i P_l \lambda_{l,l})  & \mbox{for} \ \  x<y \ , \\
		\left[\prod_{n=x}^{d} \Lambda_{y,n}  \right]\left[\prod_{m=y+1}^{d-1} \prod_{n=m+1}^{d} \Lambda_{m,n}   \right] \prod_{l=y}^{d} \exp(i P_l \lambda_{l,l}) & \mbox{for} \ \ x>y \ . \end{array}\right.
\end{align}
Here, it is important to realize that $U_{x,y}^{\dagger}Y_{x,y}U_{x,y}$ and $U_{x,y}^{\dagger}P_{x}U_{x,y}$ do no longer contain operations $\Lambda_{m,n}$ with $m< \min\{x,y\}$, meaning that there are no off-diagonal elements $\ket{m}\bra{n}$ and $\ket{n}\bra{m}$ with $m< \min\{x,y\}$. This and the observation (\ref{diagglobal}) imply that when we choose the following orthogonal operator basis and order
\begin{align}
\label{basisorder}
\begin{array}{cccccccc}
b_1&=&\ket{1}\bra{2}+\ket{2}\bra{1}\ ,& \hspace{1cm} &\alpha_1&=&\lambda_{2,1}\ ,\\
b_2&=&-i\ket{1}\bra{2}+i\ket{2}\bra{1}\ ,& \hspace{1cm} &\alpha_2&=&\lambda_{1,2}\ ,\\
b_3&=&\ket{1}\bra{3}+\ket{3}\bra{1}\ ,& \hspace{1cm} &\alpha_3&=&\lambda_{3,1}\ ,\\
b_4&=&-i\ket{1}\bra{3}+i\ket{3}\bra{1}\ ,& \hspace{1cm} &\alpha_4&=&\lambda_{1,3}\ ,\\
&\vdots& & & &\vdots\\
b_{2(d-1)-1}&=&\ket{1}\bra{d}+\ket{d}\bra{1}\ ,& \hspace{1cm} &\alpha_{2(d-1)-1}&=&\lambda_{d,1}\ ,\\
b_{2(d-1)}&=&-i\ket{1}\bra{d}+i\ket{d}\bra{1}\ ,& \hspace{1cm} &\alpha_{2(d-1)}&=&\lambda_{1,d}\ ,\\
b_{2(d-1)+1}&=&\ket{2}\bra{3}+\ket{3}\bra{2}\ ,& \hspace{1cm} &\alpha_{2(d-1)+1}&=&\lambda_{3,2}\ ,\\
b_{2(d-1)+2}&=&-i\ket{2}\bra{3}+i\ket{3}\bra{2}\ ,& \hspace{1cm} &\alpha_{2(d-1)+2}&=&\lambda_{2,3}\ ,\\
b_{2(d-1)+3}&=&\ket{2}\bra{4}+\ket{4}\bra{2}\ ,& \hspace{1cm} &\alpha_{2(d-1)+3}&=&\lambda_{4,2}\ ,\\
b_{2(d-1)+4}&=&-i\ket{2}\bra{4}+i\ket{4}\bra{2}\ ,& \hspace{1cm} &\alpha_{2(d-1)+4}&=&\lambda_{2,4}\ ,\\
&\vdots& & & &\vdots\\
b_{d^2-d-1}&=&\ket{d-1}\bra{d}+\ket{d}\bra{d-1}\ ,& \hspace{1cm} &\alpha_{d^2-d-1}&=&\lambda_{d,d-1}\ ,\\
b_{d^2-d}&=&-i\ket{d-1}\bra{d}+i\ket{d}\bra{d-1}\ ,& \hspace{1cm} &\alpha_{d^2-d}&=&\lambda_{d-1,d}\ ,\\
b_{d^2-d+1}&=&\ket{1}\bra{1}\ ,& \hspace{1cm} &\alpha_{d^2-d+1}&=&\lambda_{1,1}\ ,\\
b_{d^2-d+2}&=&\ket{2}\bra{2}\ ,& \hspace{1cm} &\alpha_{d^2-d+2}&=&\lambda_{2,2}\ ,\\
&\vdots& & & &\vdots\\
b_{d^2}&=&\ket{d}\bra{d}\ ,& \hspace{1cm} &\alpha_{d^2}&=&\lambda_{d,d}\ ,&\\
\end{array}
\end{align}
the Jacobian matrix becomes a lower block-triangular matrix\\
\setlength{\unitlength}{1cm}
\begin{picture}(14,9)(0,0)
 \put(3,4.5){\fontsize{12}{12}\selectfont\makebox(0,0)[]{$(j'_{k,l}) \ = $ \strut}}
\put(5,0){\includegraphics[scale=0.4]{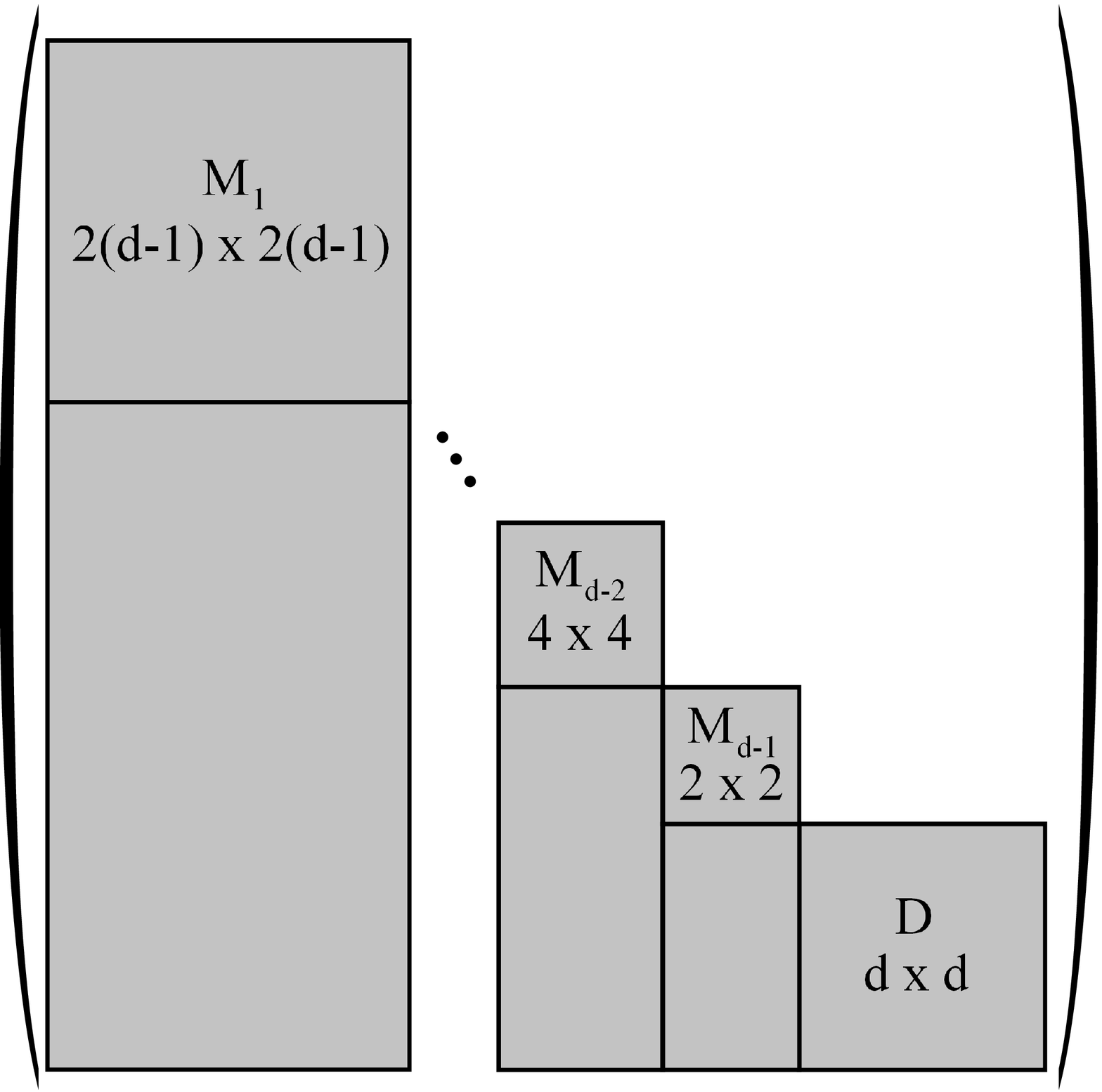}}
\end{picture}\\
where all entries outside the grey-shaded blocks are zero. Thus, the Jacobian determinant simplifies to a product of the determinants of the blocks $M_i$ and $D$
\begin{align}
J_d=\left| \det (j'_{k,l}) \right|= \left(\prod_{i=1}^{d-1}\left| \det M_i \right| \right) \left| \det D \right| \ .
\end{align}
Furthermore, we have that $\left| \det D \right|=1$ since $D$ is a $d\times d$ identity matrix due to (\ref{diagglobal}) and the choice (\ref{basisorder}). It now remains to investigate the blocks $M_i$. Let us first consider $-iU_C^{\dagger}\frac{\partial U_C}{\partial \lambda_{x,y}}$ for the Hilbert space $\mathcal{H}'=\mathbb{C}^{d-1}$ whose dimension is lower by one. In this case we have
\begin{align}
\label{discussion}
-iU_C^{\dagger}\frac{\partial U_C}{\partial \lambda_{x,y}}\, = \,\left\{\begin{array}{ll} U_{x,y}^{\dagger}Y_{x,y}U_{x,y} \hspace{1.3cm} & \mbox{for} \ \  x<y  \\
		U_{x,y}^{\dagger}P_{x}U_{x,y} & \mbox{for} \ \ x>y \ , \end{array}\right.
\end{align}
where
\begin{align}
U_{x,y}\, = \,\left\{\begin{array}{ll} \left[\prod_{n=y+1}^{d-1} \Lambda_{x,n}  \right]\left[\prod_{m=x+1}^{d-2} \prod_{n=m+1}^{d-1} \Lambda_{m,n}   \right] \prod_{l=x}^{d-1} \exp(i P_l \lambda_{l,l}) \hspace{0.3cm} & \mbox{for} \ \  x<y  \\
		\left[\prod_{n=x}^{d-1} \Lambda_{y,n}  \right]\left[\prod_{m=y+1}^{d-2} \prod_{n=m+1}^{d-1} \Lambda_{m,n}   \right] \prod_{l=y}^{d-1} \exp(i P_l \lambda_{l,l}) & \mbox{for} \ \ x>y \ . \end{array}\right.
\end{align}
By substituting $\lambda_{m,n} \rightarrow \lambda_{m+1,n+1}$, $\ket{m} \rightarrow \ket{m+1}$ and $\bra{m} \rightarrow \bra{m+1}$ one can directly see that these matrices are equal to $-iU_C^{\dagger}\frac{\partial U_C}{\partial \lambda_{x,y}}$ with $x,y\geq2$ and $U_C$ for the Hilbert space $\mathcal{H}=\mathbb{C}^{d}$ of full dimension $d$. Hence, since $J_d=\left| \det M_1  \right|\prod_{i=2}^{d-1}\left| \det M_i \right|$ we have established the recursion formula
\begin{align}
\label{recursion}
J_d=\left| \det M_1 \right|J_{d-1} \ ,
\end{align}
where the parameters in $J_{d-1}$ are to be substituted according to $\lambda_{m,n}\rightarrow\lambda_{m+1,n+1}$. This relation is a direct consequence of the fact that each $-iU_C^{\dagger}\frac{\partial U_C}{\partial \lambda_{x,y}}$ with $x,y\geq 2$ only contains parameters $\lambda_{m,n}$ with $m,n\geq 2$. As discussed in Section \ref{remarksUSU} the set of $(d-1)^2$ parameters $\lambda_{m,n}$ satisfying $m,n\geq 2$ correspond to the unitary group $\mathcal{U}(d-1)$ for the $(d-1)$-dimensional subspace $\spann(\ket{2},\ldots,\ket{d})$. Consequently, (\ref{recursion}) simply reflects the well-known fact \cite{TilmaSUN,TilmaUN} that the infinitesimal volume element $dU_d$ on $\mathcal{U}(d)$ is the product of the infinitesimal volume element $dU_{d-1}$ on $\mathcal{U}(d-1)$ times a to-be-determined function $g(d)$, i.e.
\begin{align}
\label{invrel}
dU_d=g(d)dU_{d-1} \ .
\end{align}
Now, due to the above mentioned reasons, for the composite parameterization $dU_{d-1}$ is a function of the parameters $\lambda_{m,n}$ with $m,n\geq 2$. In addition to that, g(d) is a function of the $2d-1$ parameters $\{\lambda_{1,1},\lambda_{1,2},\ldots,\lambda_{1,d}\}$ and $\{\lambda_{2,1},\ldots,\lambda_{d,1}\}$ since they extend the unitary group $\mathcal{U}(d-1)$ acting on $\spann(\ket{2},\ldots,\ket{d})$ to the entire $\mathcal{U}(d)$ on $\mathcal{H}=\mathbb{C}^{d}$, hence, $g(d)=g(d;\lambda_{1,1},\lambda_{1,2},\ldots,\lambda_{1,d};\lambda_{2,1},\ldots,\lambda_{d,1})$. For the sake of clarity, let us illustrate this by the example $dU_2 \rightarrow dU_3$ by comparing (\ref{U2}) and (\ref{U3}): $dU_2$ is (up to a constant) given by $\sin\lambda_{1,2} \cos\lambda_{1,2} d\lambda_{1,1}d\lambda_{1,2}d\lambda_{2,1}d\lambda_{2,2}$. According to our considerations $dU_3$ must contain the same expression in terms of $\lambda_{2,2},\lambda_{2,3},\lambda_{3,2},\lambda_{3,3}$ and it must look like
\begin{align}
dU_3\sim g(3;\lambda_{1,1},\lambda_{1,2},\lambda_{1,3};\lambda_{2,1},\lambda_{3,1})\sin \lambda_{2,3} \cos \lambda_{2,3} d\lambda_{2,2}d\lambda_{2,3}d\lambda_{3,2}d\lambda_{3,3} \ .
\end{align}
Equation (\ref{U3}) shows that this is indeed the case since it can easily be observed that
\begin{align*}
g(3;\lambda_{1,1},\lambda_{1,2},\lambda_{1,3};\lambda_{2,1},\lambda_{3,1}) \sim \sin \lambda_{1,2} \cos \lambda_{1,2} \sin \lambda_{1,3} \cos^3 \lambda_{1,3}  d \lambda_{1,1}d\lambda_{1,2}d\lambda_{1,3}d\lambda_{2,1}d\lambda_{3,1} \ .
\end{align*}
It now remains to find a general expression for $\left| \det M_1 \right|$ for arbitrary $d$, since the formula of $J_d$ can then be derive via (\ref{recursion}) and induction. In general, the matrix $M_1$ does not have a simple form which makes it difficult to compute its determinant. Here, it is useful to exploit that, first, there is no explicit dependence of $J_d$ on parameters corresponding to phase operations, i.e. all $\lambda_{m,n}$ with $m\geq n$. This can directly be followed from the construction of the composite parameterization in combination with the Haar measure on $\mathcal{U}(2)$ and the Jacobian matrix (\ref{jacobianU2}). Since in the Jacobian matrix (\ref{jacobianU2}) and in its determinant the phase operations $\lambda_{1,1},\lambda_{2,1},\lambda_{2,2}$ only appear as $e^{i\lambda_{1,1}},e^{i\lambda_{2,1}},e^{i\lambda_{2,2}}$, i.e. complex numbers of magnitude 1, and since in the end we are only interested in the absolute value of this determinant, it is clear that $dU_2$ (\ref{U2}) is independent of these parameters. As the composite parameterization only combines $\mathcal{U}(2)$-operations by incorporating all possible $2$-dimensional subspaces in $\mathbb{C}^d$ the same statement holds of course true for all $d$ (see for instance (\ref{U3}) for the case $d=3$). Second, the circumstance that $g(d)$ is independent of all parameters $\lambda_{m,n}$ with $\min\{m,n\}\geq2$ implies that these parameters do not affect $\left| \det M_1 \right|$\footnote{The independence of $\left| \det M_1 \right|$ on $\lambda_{m,n}$ with $\min\{m,n\}\geq2$ can also be confirmed by exploiting again the left and right invariance of the Haar measure.}. For these reasons one can set the parameters $\lambda_{m,n}$ with $\min\{m,n\}\geq2$ and $m\geq n$ to zero in (\ref{drotU}) without altering $|\det M_1|$, i.e.
\begin{align}
\label{Moverline}
|\det M_1|=\left|\det \left(\left. M_1 \right|_{\{\lambda_{m,n}=0|(m,n) \notin \{(1,2),\ldots,(1,d) \} \}}\right)\right| \equiv |\det \overline{M}_1| \ .
\end{align}
Here, (\ref{derivs}) reduces to
\begin{align*}
\left.U_{1,y}^{\dagger}Y_{1,y}U_{1,y}\right|_{\{\lambda_{m,n}=0|(m,n) \notin \{(1,2),\ldots,(1,d) \} \}}=&\left[ \prod_{n=y+1}^{d} \exp(i Y_{1,n} \lambda_{1,n})\right]^{\dagger} Y_{1,y} \left[ \prod_{n=y+1}^{d} \exp(i Y_{1,n} \lambda_{1,n})\right] \ , \\
\equiv & O_{1,y}^{T}Y_{1,y} O_{1,y}\\
\left.U_{x,1}^{\dagger}P_{x}U_{x,1}\right|_{\{\lambda_{m,n}=0|(m,n) \notin \{(1,2),\ldots,(1,d) \} \}} =&\left[\prod_{n=x}^{d} \exp(i Y_{1,n} \lambda_{1,n})\right]^{\dagger} P_{x} \left[\prod_{n=x}^{d} \exp(i Y_{1,n} \lambda_{1,n})\right] \ ,\\
\equiv & O_{x,1}^{T}P_{x}O_{x,1} \ ,
\end{align*}
where $O_{x,1}$ and $O_{1,y}$ are orthogonal (real) matrices since they are a product of the operations $\exp(i Y_{1,n} \lambda_{1,n})$ which explicitly read
\begin{align}
\label{explicit}
\cos (\lambda_{1,n})\ket{1}\bra{1} + \sin (\lambda_{1,n})\ket{1}\bra{n}-\sin(\lambda_{1,n})\ket{n}\bra{1}+ \cos (\lambda_{m,n})\ket{n}\bra{n} + \sum_{k \neq 1,n}& \ket{k}\bra{k} \ .
\end{align}
According to (\ref{basisorder}), the $2(d-1) \times 2 (d-1)$ elements of $\overline{M}_1$ are now determined by
\begin{align}
\label{Yexp}
\Tr\left(b^{\dagger}_k O_{1,y}^{T}Y_{1,y}O_{1,y}\right)=\Tr\left(O_{1,y}b^{\dagger}_k O_{1,y}^{T}Y_{1,y}\right)
\end{align}
and
\begin{align}
\label{Pexp}
\Tr\left(b^{\dagger}_k O_{x,1}^{T}P_{x}O_{x,1}\right)=\Tr\left(O_{x,1}b^{\dagger}_k O_{x,1}^{T}P_{x}\right)
\end{align}
with $x,y\in \{2,\ldots,d\}$, and
\begin{align}
\label{m1basis}
b_k=b^{\dagger}_k\, = \,\left\{\begin{array}{lll} \ket{1}\bra{(k+3)/2}+\ket{(k+3)/2}\bra{1} &= \ X_{1,(k+3)/2} \hspace{0.3cm} & k \ \mbox{- odd} \ ,\\
		-i\ket{1}\bra{(k+2)/2}+i\ket{(k+2)/2}\bra{1} &= \ Y_{1,(k+2)/2} & k \ \mbox{- even} \ , \end{array}\right.
\end{align}
with $k\in\{1,\ldots,2(d-1)\}$. Now consider the coefficients (\ref{Yexp}) and (\ref{Pexp}) depending on different $X_{1,m}$ and $Y_{1,m} \ (2 \leq m \leq d)$ of (\ref{m1basis}):\\
\vspace{-0.3cm}\\
$\bullet$ $\Tr\left(O_{1,y}X_{1,m} O_{1,y}^{T}Y_{1,y}\right)$ for arbitrary $m$:
\begin{align}
\Tr\left(O_{1,y}X_{1,m} O_{1,y}^{T}Y_{1,y}\right)&=0 \ .
\end{align}
\hspace{0.2cm}Note: $O_{1,y}X_{1,m} O_{1,y}^{T}$ is symmetric while $Y_{1,y}$ is antisymmetric.\\
\vspace{-0.3cm}\\$\bullet$ $\Tr\left(O_{1,y}Y_{1,m} O_{1,y}^{T}Y_{1,y}\right)$ for $m < y$:
\begin{align}
\Tr\left(O_{1,y}Y_{1,m} O_{1,y}^{T}Y_{1,y}\right)=0 \ .
\end{align}
\hspace{0.2cm}Note: $O_{1,y}Y_{1,m} O_{1,y}^{T}$ is orthogonal to $Y_{1,y}$ for $m < y$ since the product in\\
$O_{1,y}=\prod_{n=y+1}^{d} \exp(i Y_{1,n} \lambda_{1,n})$ starts with $n=y+1$.\\
\vspace{-0.3cm}\\
$\bullet$ $\Tr\left(O_{x,1}Y_{1,m} O_{x,1}^{T}P_{x}\right)$ for arbitrary $m$:
\begin{align}
\Tr\left(O_{x,1}Y_{1,m} O_{x,1}^{T}P_{x}\right)&=\bra{x}O_{x,1}Y_{1,m} O_{x,1}^{T}\ket{x} \ ,\\
&=-i\bra{x}O_{x,1}\ket{1}\bra{m} O_{x,1}^{T}\ket{x}+i\bra{x}O_{x,1}\ket{m}\bra{1} O_{x,1}^{T}\ket{x} \ ,\\
&=0 \ .
\end{align}
\hspace{0.2cm}Note: $O_{x,1}$ is an orthogonal (real) matrix.\\
\vspace{-0.3cm}\\
$\bullet$ $\Tr\left(O_{x,1}X_{1,m} O_{x,1}^{T}P_{x}\right)$ for $m < x$:
\begin{align}
\Tr\left(O_{x,1}X_{1,m} O_{x,1}^{T}P_{x}\right)&=\bra{x}O_{x,1}X_{1,m} O_{x,1}^{T}\ket{x} \ , \\
&=\bra{x}O_{x,1}\ket{1}\bra{m} O_{x,1}^{T}\ket{x}+\bra{x}O_{x,1}\ket{m}\bra{1} O_{x,1}^{T}\ket{x} \ ,\\
&=0 \ .
\end{align}
\hspace{0.2cm}Note: $O_{x,1}=\prod_{n=x}^{d} \exp(i Y_{1,n} \lambda_{1,n})$ has no off-diagonal elements $\bra{x}O_{x,1}\ket{m}$ for $m<x$.\\
\vspace{-0.3cm}\\
These four observations imply for the operator basis and parameter order
\begin{align}
\begin{array}{cccccccc}
b_1&=&\ket{1}\bra{2}+\ket{2}\bra{1}\ ,& \hspace{1cm} &\alpha_1&=&\lambda_{2,1}\ ,\\
b_2&=&-i\ket{1}\bra{2}+i\ket{2}\bra{1}\ ,& \hspace{1cm} &\alpha_2&=&\lambda_{1,2}\ ,\\
b_3&=&\ket{1}\bra{3}+\ket{3}\bra{1}\ ,& \hspace{1cm} &\alpha_3&=&\lambda_{3,1}\ ,\\
b_4&=&-i\ket{1}\bra{3}+i\ket{3}\bra{1}\ ,& \hspace{1cm} &\alpha_4&=&\lambda_{1,3}\ ,\\
&\vdots& & & &\vdots\\
b_{2(d-1)-1}&=&\ket{1}\bra{d}+\ket{d}\bra{1}\ ,& \hspace{1cm} &\alpha_{2(d-1)-1}&=&\lambda_{d,1}\ ,\\
b_{2(d-1)}&=&-i\ket{1}\bra{d}+i\ket{d}\bra{1}\ ,& \hspace{1cm} &\alpha_{2(d-1)}&=&\lambda_{1,d}\ ,\\
\end{array}
\end{align}
that $\overline{M}_1$ is a lower triangular matrix\\
\setlength{\unitlength}{1cm}
\begin{picture}(14,5)(0,0)
 \put(5,2){\fontsize{12}{12}\selectfont\makebox(0,0)[]{$\overline{M}_1= (j'_{k,l})_{\mbox{\tiny{$k,l=1,\ldots,2(d-1)$}}} =$ \strut}}
\put(8.5,0){\includegraphics[scale=0.2]{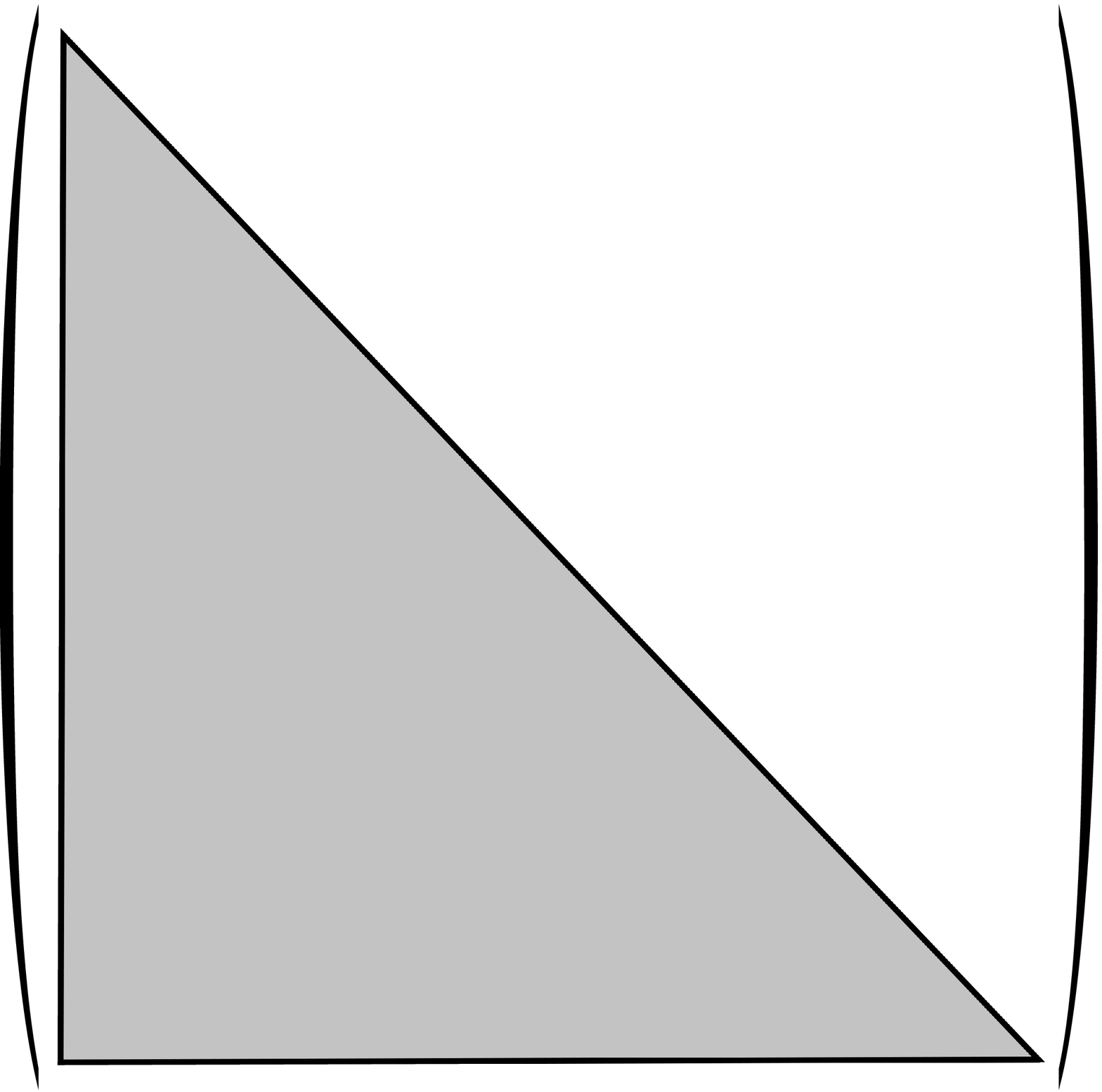}}
\put(13.5,0.4){\fontsize{12}{12}\selectfont\makebox(0,0)[]{$.$ \strut}}
\end{picture}\\
For the determinant we can now restrict on determining the diagonal entries of $\overline{M}_1$ which are given by\\
\vspace{-0.3cm}\\
$\bullet$ $\Tr\left(O_{1,y}Y_{1,m} O_{1,y}^{T}Y_{1,y}\right)/\Tr(Y^2_{1,m})$ for $m = y$:
\begin{align}
\Tr\left(O_{1,m}Y_{1,m} O_{1,m}^{T}Y_{1,m}\right)=&\bra{1}O_{1,m}Y_{1,m} O_{1,m}^{T}Y_{1,m}\ket{1}+\bra{m}O_{1,m}Y_{1,m} O_{1,m}^{T}Y_{1,m}\ket{m} \ ,\\
=&i\bra{1}O_{1,m}Y_{1,m} O_{1,m}^{T}\ket{m}-i\bra{m}O_{1,m}Y_{1,m} O_{1,m}^{T}\ket{1} \ , \\
=&\bra{1}O_{1,m}\ket{1}\bra{m} O_{1,m}^{T}\ket{m}-\bra{1}O_{1,m}\ket{m}\bra{1} O_{1,m}^{T}\ket{m} \nonumber \\
 &-\bra{m}O_{1,m}\ket{1}\bra{m} O_{1,m}^{T}\ket{1}+\bra{m}O_{1,m}\ket{m}\bra{1} O_{1,m}^{T}\ket{1} \ , \\
=&2\bra{1}O_{1,m}\ket{1}\bra{m} O_{1,m}\ket{m} \ . \label{coeff2}
\end{align}
\hspace{0.2cm}Note: $O_{1,m}=\prod_{n=m+1}^{d} \exp(i Y_{1,n} \lambda_{1,n})$ does not have off-diagonal element $\bra{1}O_{1,m}\ket{m}$ since the product starts with $n=m+1$.\\
$\bullet$ $\Tr\left(O_{x,1}X_{1,m} O_{x,1}^{T}P_{x}\right)/\Tr(X^2_{1,m})$ for $m = x$:
\begin{align}
\Tr\left(O_{m,1}X_{1,m} O_{m,1}^{T}P_{m}\right)&=\bra{m}O_{m,1}X_{1,m} O_{m,1}^{T}\ket{m} \ , \\
&=\bra{m}O_{m,1}\ket{1}\bra{m}O_{m,1}^{T}\ket{m}+\bra{m}O_{m,1}\ket{m}\bra{1} O_{m,1}^{T}\ket{m} \ , \\
&=2\bra{m}O_{m,1}\ket{1}\bra{m}O_{m,1}\ket{m} \ . \label{coeff1}
\end{align}
\vspace{-0.3cm}\\
We have thus found simple relations between the diagonal entries of $\overline{M}_1$ and the matrix elements ($2\leq m \leq d$)
\begin{align}
\bra{1}O_{1,m}\ket{1}&=\prod_{n=m+1}^{d}\cos \lambda_{1,n} \ , \\
\bra{m} O_{1,m}\ket{m}&=1  \ , \\
\label{rel1}
\bra{m}O_{m,1}\ket{1}&=-\sin \lambda_{1,m}\prod_{n=m+1}^{d}\cos \lambda_{1,n} \ , \\
\label{rel2}
\bra{m}O_{m,1}\ket{m}&=\cos \lambda_{1,m}  \ ,
\end{align}
which can easily be obtained via the definitions of $O_{m,1}$ and $O_{1,m}$ together with (\ref{explicit}). Hence, since $\Tr(X^2_{1,m})=\Tr(Y^2_{1,m})=2$, we find that the diagonal entries of $\overline{M}_1$ are
\begin{align}
\label{diagelements1}
\Tr\left(O_{1,m}Y_{1,m} O_{1,m}^{T}Y_{1,m}\right)/\Tr(Y^2_{1,m})&=\prod_{n=m+1}^{d}\cos \lambda_{1,n}  \ , \\
\label{diagelements2}
\Tr\left(O_{m,1}X_{1,m} O_{m,1}^{T}P_{m}\right)/\Tr(X^2_{1,m})&=-\sin \lambda_{1,m}\prod_{n=m}^{d}\cos \lambda_{1,n} \ .
\end{align}
If we multiply all these entries ($m=2,\ldots,d$) we finally obtain
\begin{align}
\left|\det \overline{M}_1 \right|&= \left[\prod_{k=2}^{d}\left(\prod_{n=k+1}^{d}\cos \lambda_{1,n}\right)\right] \cdot \left[\prod_{l=2}^{d}\left( \sin \lambda_{1,l}\prod_{n=l}^{d}\cos \lambda_{1,n}\right) \right] \ , \\
&=\prod_{n=2}^{d}\sin(\lambda_{1,n})\cos^{2(n-1)-1}(\lambda_{1,n}) \ .
\end{align}
Using the recursion formula (\ref{recursion}) and induction one finds the final result
\begin{align}
\label{finalJd}
J_d=\prod_{m=1}^{d-1}\prod_{n=m+1}^{d}\sin(\lambda_{m,n})\cos^{2(n-m)-1}(\lambda_{m,n}) \ .
\end{align}
The corresponding normalizing constant $N_d$ is given by the integral
\begin{align}
N_d=&\int_{\mathcal{U}(d)}J_d\prod_{k,l=1}^{d}d\lambda_{k,l} \ , \\
=&\int_{\lambda_{k,l}=0}^{2\pi \ (k\geq l)}\int_{\lambda_{k,l}=0}^{\pi/2 \ (k<l)} \left[\prod_{m=1}^{d-1}\prod_{n=m+1}^{d}\sin(\lambda_{m,n})\cos^{2(n-m)-1}(\lambda_{m,n}) \right] \prod_{k,l=1}^{d}d\lambda_{k,l} \ ,
\end{align}
which can straightforwardly be computed
\begin{align}
N_d=&(2\pi)^{d(d+1)/2} \prod_{m=1}^{d-1}\prod_{n=m+1}^{d} \left. \frac{-1}{2(n-m)}\cos^{2(n-m)}(\lambda_{m,n}) \right|_{\lambda_{m,n}=0}^{\pi/2} \ , \\
=&\frac{(2\pi)^{d(d+1)/2}}{\prod_{m=1}^{d-1}\prod_{n=m+1}^{d} 2(n-m)} \ .
\end{align}

\end{proof}
\section{Haar measure on the special unitary group $\mathcal{SU}(d)$}
\label{haarSU}
\begin{samepage}
\begin{thm}
\label{theorem4}
In terms of the $d^2-1$ parameters $\lambda_{m,n}$ introduced in Theorem $2$ the (normalized) Haar measure on the special unitary group $\mathcal{SU}(d)$ reads
\begin{align}
\label{HaarSU}
dU_{d}=\frac{1}{N_d}\prod_{m=1}^{d-1}\prod_{n=m+1}^{d}\sin(\lambda_{m,n})\cos^{2(n-m)-1}(\lambda_{m,n}) \prod_{k,l}d\lambda_{k,l} \ ,
\end{align}
with
\begin{align}
N_d=\frac{2^{d-1}\pi^{d(d+1)/2-1}}{\prod_{m=1}^{d-1}\prod_{n=m+1}^{d} 2(n-m)}
\end{align}
such that $\int_{\mathcal{SU}(d)} dU_{d} =1$.
\end{thm}
\end{samepage}
\begin{proof}
The proof is similar to the previous one --- only minor modifications have to be made. Given the special unitary group $\mathcal{SU}(d)$ in parameterized form $U(\alpha_1,\ldots,\alpha_{d^2-1})$, to construct the Haar measure one must determine the absolute value $J_d = \left| \det (j_{k,l}) \right|$ of the determinant of the Jacobian matrix
\begin{align}
(j_{k,l})=\frac{\partial(u_1,\ldots,u_{d^2-1})}{\partial (\alpha_1,\ldots,\alpha_{d^2-1})} \ ,
\end{align}
wherein $\{ u_k \}$ are coefficients of the special unitary operator $U(\alpha_1,\ldots,\alpha_{d^2-1})$ expanded in an orthogonal basis $\{ b_k \}$ of traceless operators\footnote{A basis for the vector space of operators $\mathbb{C}^d \times \mathbb{C}^d$ has $d^2$ elements. The constraint $\det U=1$ on special unitary operators implies that the trace of any derivative $\partial U / \partial \alpha_l$ with respect to an arbitrary parameter is always zero. Traceless operators form a $d^2-1$ dimensional subspace of $\mathbb{C}^d \times \mathbb{C}^d$. As we have $d^2-1$ parameters $\{ \alpha_l \}$ it is required to express the derivatives
$\partial U / \partial \alpha_l$ in a basis of this subspace to obtain $d^2-1$ linearly independent column vectors.} of $\mathbb{C}^d \times \mathbb{C}^d$, i.e.
\begin{align}
u_k=\frac{\Tr(b_k^{\dagger}U)}{\Tr(b_k^{\dagger}b_k)} \hspace{1.8cm} \left(  \Rightarrow \frac{\partial U}{\partial \alpha_l}=\sum_{k=1}^{d^2-1} \frac{\partial u_k}{\partial \alpha_l} b_k  \right) \ .
\end{align}
In this terminology, the (normalized) Haar measure reads
\begin{align}
dU_d=\frac{J_d}{N_d} \prod_{l=1}^{d^2-1} d\alpha_l \ .
\end{align}
Our aim is to derive a general expression of $dU_d$ for arbitrary $d$ in terms of the parameterization introduced in Section \ref{compSU}, i.e.
\begin{align}
dU_d=\frac{J_d}{N_d} \prod_{k,l}d\lambda_{k,l} \ .
\end{align}
As in the previous proof we make use of the left invariance of the Haar measure on $\mathcal{SU}(d)$, i.e. $J_d=|\det (j_{k,l})|=|\det (j'_{k,l})|$ where
\begin{align}
j_{k,l}=\frac{\Tr(b_k^{\dagger} \frac{\partial U}{\partial \alpha_l})}{\Tr(b_k^{\dagger}b_k)} \hspace{0.5cm} \longrightarrow \hspace{0.5cm} j'_{k,l}=\frac{\Tr(b_k^{\dagger} U_1 \frac{\partial U}{\partial \alpha_l})}{\Tr(b_k^{\dagger}b_k)} \ .
\end{align}
If for the composite parameterization (Theorem \ref{theorem2}) $U_1$ is chosen to be $-iU^{\dagger}_C$ one obtains analogously to (\ref{diagglobal}) -- (\ref{drotU}) that
\begin{align}
\label{derivs2}
-iU_C^{\dagger}\frac{\partial U_C}{\partial \lambda_{x,y}}\, = \,\left\{\begin{array}{ll} U_{x,y}^{\dagger}Y_{x,y}U_{x,y} \hspace{1cm}  & \mbox{for} \ \  x<y  \\
		\hspace{0.7cm} Z_{x,d}  & \mbox{for} \ \ x=y \  \\
U_{x,y}^{\dagger}Z_{y,x}U_{x,y} & \mbox{for} \ \ x>y \ , \end{array}\right.
\end{align}
where
\begin{align}
U_{x,y}\, = \,\left\{\begin{array}{ll} \left[\prod_{n=y+1}^{d} \Lambda_{x,n}  \right]\left[\prod_{m=x+1}^{d-1} \prod_{n=m+1}^{d} \Lambda_{m,n}   \right] \prod_{l=x}^{d-1} \exp(i Z_{l,d} \lambda_{l,l}) & \mbox{for} \ \  x<y  \\
		\left[\prod_{n=x}^{d} \Lambda_{y,n}  \right]\left[\prod_{m=y+1}^{d-1} \prod_{n=m+1}^{d} \Lambda_{m,n}   \right] \prod_{l=y}^{d-1} \exp(i Z_{l,d} \lambda_{l,l}) & \mbox{for} \ \ x>y \ . \end{array}\right.
\end{align}
These operators are now to be expressed in an orthogonal basis $\{ b_k \}$ of traceless operators. Since the first $d^2-d$ operators that were used in (\ref{basisorder}) already are mutually orthogonal and traceless, only the $d$ diagonal operators $\ket{k}\bra{k}$ with $1\leq k \leq d$ have to be replaced by $d-1$ diagonal operators with vanishing trace. A convenient choice are the $d-1$ mutually orthogonal operators\footnote{Note that these are the generalized diagonal Gell-Mann matrices \cite{Krammer} in reversed order.}
\begin{align}
L_k=\sqrt{\frac{2}{(d-k)(d-k+1)}} \left(  -(d-k)\ket{k}\bra{k} + \sum_{n=k+1}^{d} \ket{n} \bra{n} \right) \ ,
\end{align}
with $1\leq k \leq d-1$ obeying $\Tr(L_k^{\dagger} L_k)=\Tr(L_k^2)=2$. For the order
\begin{align}
\label{basisorderSU}
\begin{array}{cccccccc}
b_1&=&\ket{1}\bra{2}+\ket{2}\bra{1}\ ,& &\alpha_1&=&\lambda_{2,1}\ ,\\
b_2&=&-i\ket{1}\bra{2}+i\ket{2}\bra{1}\ ,&  &\alpha_2&=&\lambda_{1,2}\ ,\\
b_3&=&\ket{1}\bra{3}+\ket{3}\bra{1}\ ,&  &\alpha_3&=&\lambda_{3,1}\ ,\\
b_4&=&-i\ket{1}\bra{3}+i\ket{3}\bra{1}\ ,&  &\alpha_4&=&\lambda_{1,3}\ ,\\
&\vdots& & & &\vdots\\
b_{2(d-1)-1}&=&\ket{1}\bra{d}+\ket{d}\bra{1}\ ,&  &\alpha_{2(d-1)-1}&=&\lambda_{d,1}\ ,\\
b_{2(d-1)}&=&-i\ket{1}\bra{d}+i\ket{d}\bra{1}\ ,&  &\alpha_{2(d-1)}&=&\lambda_{1,d}\ ,\\
b_{2(d-1)+1}&=&\ket{2}\bra{3}+\ket{3}\bra{2}\ ,& &\alpha_{2(d-1)+1}&=&\lambda_{3,2}\ ,\\
b_{2(d-1)+2}&=&-i\ket{2}\bra{3}+i\ket{3}\bra{2}\ ,& &\alpha_{2(d-1)+2}&=&\lambda_{2,3}\ ,\\
b_{2(d-1)+3}&=&\ket{2}\bra{4}+\ket{4}\bra{2}\ ,&  &\alpha_{2(d-1)+3}&=&\lambda_{4,2}\ ,\\
b_{2(d-1)+4}&=&-i\ket{2}\bra{4}+i\ket{4}\bra{2}\ ,&  &\alpha_{2(d-1)+4}&=&\lambda_{2,4}\ ,\\
&\vdots& & & &\vdots\\
b_{d^2-d-1}&=&\ket{d-1}\bra{d}+\ket{d}\bra{d-1}\ ,& &\alpha_{d^2-d-1}&=&\lambda_{d,d-1}\ ,\\
b_{d^2-d}&=&-i\ket{d-1}\bra{d}+i\ket{d}\bra{d-1}\ ,& &\alpha_{d^2-d}&=&\lambda_{d-1,d}\ ,\\
b_{d^2-d+1}&=&L_1\ ,& \hspace{1cm} &\alpha_{d^2-d+1}&=&\lambda_{1,1}\ ,\\
b_{d^2-d+2}&=&L_2\ ,&  &\alpha_{d^2-d+2}&=&\lambda_{2,2}\ ,\\
&\vdots& & & &\vdots\\
b_{d^2-1}&=& L_{d-1}\ ,&  &\alpha_{d^2-1}&=&\lambda_{d-1,d-1}\ ,& \\
\end{array}
\end{align}
the $(d^2-1) \times (d^2-1)$ Jacobian matrix is again lower block-triangular\\
\setlength{\unitlength}{1cm}
\begin{picture}(14,9)(0,0)
 \put(3,4.5){\fontsize{12}{12}\selectfont\makebox(0,0)[]{$(j'_{k,l}) \ = $ \strut}}
\put(5,0){\includegraphics[scale=0.4]{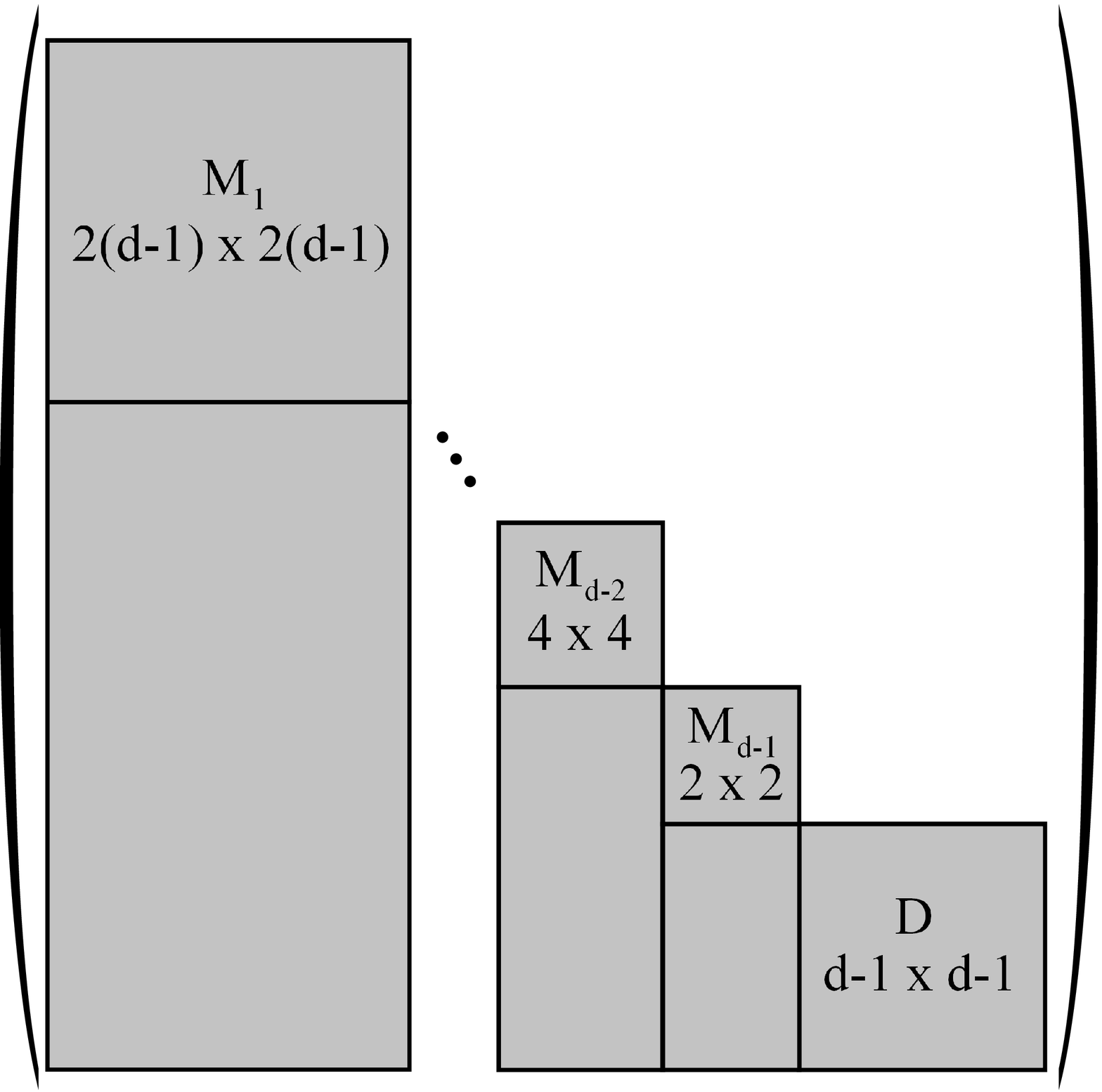}}
\put(14,0){\fontsize{12}{12}\selectfont\makebox(0,0)[]{$.$ \strut}}
\end{picture}\\
Here, the block $D$ is not diagonal but lower triangular since
\begin{align}
\Tr(L_k^{\dagger}Z_{m,d})=\,\left\{\begin{array}{ll} 0 \hspace{1cm}  & \mbox{for} \ \  k<m  \\
		\sqrt{\frac{2}{(d-k)(d-k+1)}} (k-d-1)  & \mbox{for} \ \ k=m \  \\
-\sqrt{\frac{2}{(d-k)(d-k+1)}} & \mbox{for} \ \ k>m \ . \end{array}\right.
\end{align}
Since the diagonal entries of $D$ are merely real numbers we find again that
\begin{align}
J_d=\left| \det (j'_{k,l}) \right|= c_d \prod_{i=1}^{d-1}\left| \det M_i \right|  \ ,
\end{align}
where $c_d$ is a constant that can be dropped since the Haar measure will be normalized at the end anyhow. As the composite parameterization of the unitary group and the special unitary group share the same structure the discussion between (\ref{recursion}) -- (\ref{Moverline}) essentially remains the same, meaning that (ignoring irrelevant constants) one finds again the recursion formula
\begin{align}
J_d=\left| \det \overline{M}_1 \right|J_{d-1} \ ,
\end{align}
where $ \overline{M}_1=\left. M_1 \right|_{\{\lambda_{m,n}=0|(m,n) \notin \{(1,2),\ldots,(1,d) \} \}}$ and $\lambda_{m,n} \rightarrow \lambda_{m+1,n+1}$ in $J_{d-1}$. Here, the relevant operators of (\ref{derivs2}) have the form
\begin{align*}
\left.U_{1,y}^{\dagger}Y_{1,y}U_{1,y}\right|_{\{\lambda_{m,n}=0|(m,n) \notin \{(1,2),\ldots,(1,d) \} \}}&=\left[ \prod_{n=y+1}^{d} \exp(i Y_{1,n} \lambda_{1,n})\right]^{\dagger} Y_{1,y} \left[ \prod_{n=y+1}^{d} \exp(i Y_{1,n} \lambda_{1,n})\right] \ , \\
&\equiv O_{1,y}^{T}Y_{1,y} O_{1,y} \ , \\
\left.U_{x,1}^{\dagger}Z_{1,x}U_{x,1}\right|_{\{\lambda_{m,n}=0|(m,n) \notin \{(1,2),\ldots,(1,d) \} \}}&=\left[\prod_{n=x}^{d} \exp(i Y_{1,n} \lambda_{1,n})\right]^{\dagger} Z_{1,x} \left[\prod_{n=x}^{d} \exp(i Y_{1,n} \lambda_{1,n})\right] \ , \\
&\equiv O_{x,1}^{T}Z_{1,x}O_{x,1} \ ,
\end{align*}
containing the orthogonal matrices $O_{x,1}$ and $O_{1,y}$ that also appeared in the previous proof. The $2(d-1) \times 2 (d-1)$ elements of $\overline{M}_1$ are determined by (compare with the order (\ref{basisorderSU}))
\begin{align}
\label{Yexp2}
\Tr\left(b^{\dagger}_k O_{1,y}^{T}Y_{1,y}O_{1,y}\right)=\Tr\left(O_{1,y}b^{\dagger}_k O_{1,y}^{T}Y_{1,y}\right)
\end{align}
and
\begin{align}
\label{Zexp2}
\Tr\left(b^{\dagger}_k O_{x,1}^{T}Z_{1,x}O_{x,1}\right)=\Tr\left(O_{x,1}b^{\dagger}_k O_{x,1}^{T}Z_{1,x}\right)
\end{align}
with $x,y\in \{2,\ldots,d\}$, and
\begin{align}
\label{m1blocko}
b_k=b^{\dagger}_k\, = \,\left\{\begin{array}{lll} \ket{1}\bra{(k+3)/2}+\ket{(k+3)/2}\bra{1} &= \ X_{1,(k+3)/2} \hspace{0.3cm} & k \ \mbox{- odd} \\
		-i\ket{1}\bra{(k+2)/2}+i\ket{(k+2)/2}\bra{1} &= \ Y_{1,(k+2)/2} & k \ \mbox{- even} \ , \end{array}\right.
\end{align}
with $k\in\{1,\ldots,2(d-1)\}$. Now consider the coefficients of (\ref{Yexp2}) and (\ref{Zexp2}) for the different operator basis elements occurring in (\ref{m1blocko}): First, notice that $\overline{M}_1$ is again lower triangular as in the previous proof since (\ref{Yexp2}) and (\ref{Yexp}) are identical and\\
\vspace{-0.3cm}\\
$\bullet$ $\Tr\left(O_{x,1}Y_{1,m} O_{x,1}^{T}Z_{1,x}\right)=0$ for arbitrary $m$.\\
\vspace{-0.3cm}\\
\hspace{0.2cm}Note: $O_{1,y}Y_{1,m} O_{1,y}^{T}$ is antisymmetric, while $Z_{1,x}$ is symmetric.\\
\vspace{-0.3cm}\\
$\bullet$ $\Tr\left(O_{x,1}X_{1,m} O_{x,1}^{T}Z_{1,x}\right)=0$ for $m < x$:
\begin{align}
\Tr\left(O_{x,1}X_{1,m} O_{x,1}^{T}Z_{1,x}\right)=&\bra{1}O_{x,1}X_{1,m} O_{x,1}^{T}\ket{1}-\bra{x}O_{x,1}X_{1,m} O_{x,1}^{T}\ket{x} \ , \\
=&\bra{1}O_{x,1}\ket{1}\bra{m} O_{x,1}^{T}\ket{1}-\bra{1}O_{x,1}\ket{m}\bra{1} O_{x,1}^{T}\ket{1} \nonumber\\
&-\bra{x}O_{x,1}\ket{1}\bra{m} O_{x,1}^{T}\ket{x}-\bra{x}O_{x,1}\ket{m}\bra{1} O_{x,1}^{T}\ket{x} \ , \\
=&0 \ .
\end{align}
\hspace{0.2cm}Note: $O_{x,1}=\prod_{n=x}^{d} \exp(i Y_{1,n} \lambda_{1,n})$ does not have off-diagonal elements $\bra{1}O_{x,1}\ket{m}$ and $\bra{x}O_{x,1}\ket{m}$ for $m<x$.\\
\vspace{-0.3cm}\\
Thus, it again suffices to compute the diagonal entries of $\overline{M}_1$, half of which are already known from (\ref{diagelements1})
\begin{align}
\label{diag1}
\Tr\left(O_{1,m}Y_{1,m} O_{1,m}^{T}Y_{1,m}\right)/\Tr(Y^2_{1,m})&=\prod_{n=m+1}^{d}\cos \lambda_{1,n} \ .
\end{align}
The remaining ones are found to be
\begin{align}
\label{diag2}
\Tr\left(O_{1,m}X_{1,m} O_{1,m}^{T}Z_{1,m}\right)/\Tr(X^2_{1,m})&=2\sin \lambda_{1,m}\prod_{n=m}^{d}\cos \lambda_{1,n} \ ,
\end{align}
which directly follow from
\begin{align}
\Tr\left(O_{1,m}X_{1,m} O_{1,m}^{T}Z_{1,m}\right)=&\bra{1}O_{m,1}X_{1,m} O_{m,1}^{T}\ket{1}-\bra{m}O_{m,1}X_{1,m} O_{m,1}^{T}\ket{m} \ , \\
=&\bra{1}O_{m,1}\ket{1}\bra{m}O_{m,1}^{T}\ket{1}+\bra{1}O_{m,1}\ket{m}\bra{1} O_{m,1}^{T}\ket{1} \nonumber\\
 &-\bra{m}O_{m,1}\ket{1}\bra{m}O_{m,1}^{T}\ket{m}-\bra{m}O_{m,1}\ket{m}\bra{1} O_{m,1}^{T}\ket{m} \ , \nonumber\\
=&2\bra{1}O_{m,1}\ket{m}\bra{1}O_{m,1}\ket{1}-2\bra{m}O_{m,1}\ket{1}\bra{m}O_{m,1}\ket{m} \label{coeffsu1} \ .
\end{align}
and the definition of $O_{m,1}$ together with (\ref{explicit})
\begin{align}
\bra{1}O_{m,1}\ket{m}&=\sin \lambda_{1,m} \ , \\
\bra{1}O_{m,1}\ket{1}&=\prod_{n=m}^{d}\cos \lambda_{1,n}  \ , \\
\bra{m}O_{m,1}\ket{1}&=-\sin \lambda_{1,m}\prod_{n=m+1}^{d}\cos \lambda_{1,n} \ , \\
\bra{m}O_{m,1}\ket{m}&=\cos \lambda_{1,m} \ .
\end{align}
Since (\ref{diag2}) differs only by the factor two from (\ref{diagelements1}) we again obtain for $J_d$ the result (\ref{finalJd})
\begin{align}
J_d=c'_d \prod_{m=1}^{d-1}\prod_{n=m+1}^{d}\sin(\lambda_{m,n})\cos^{2(n-m)-1}(\lambda_{m,n}) \ ,
\end{align}
up to an irrelevant multiplicative constant $c'_d$. Hence, the Haar measure on $\mathcal{SU}(d)$ reads
\begin{align}
dU_{d}=\frac{1}{N_d}\prod_{m=1}^{d-1}\prod_{n=m+1}^{d}\sin(\lambda_{m,n})\cos^{2(n-m)-1}(\lambda_{m,n}) \prod_{k,l}d\lambda_{k,l} \ .
\end{align}
The normalizing constant $N_d$ is given by the integral
\begin{align}
N_d=&\int_{\lambda_{k,l}=0}^{2\pi \ (k=l)}\int_{\lambda_{k,l}=0}^{\pi \ (k> l)}\int_{\lambda_{k,l}=0}^{\pi/2 \ (k<l)}  \left[\prod_{m=1}^{d-1}\prod_{n=m+1}^{d}\sin(\lambda_{m,n})\cos^{2(n-m)-1}(\lambda_{m,n}) \right] \prod_{k,l}d\lambda_{k,l} \ . \nonumber
\end{align}
Since there are $d-1$ parameters $\lambda_{k,l}$ with $k=l$ and $d(d-1)/2$ parameters $\lambda_{k,l}$ with $k>l$ one finally obtains
\begin{align}
N_d=&2^{d-1}\pi^{d(d+1)/2-1} \prod_{m=1}^{d-1}\prod_{n=m+1}^{d} \left. \frac{-1}{2(n-m)}\cos^{2(n-m)}(\lambda_{m,n}) \right|_{\lambda_{m,n}=0}^{\pi/2} \ ,  \\
=&\frac{2^{d-1}\pi^{d(d+1)/2-1}}{\prod_{m=1}^{d-1}\prod_{n=m+1}^{d} 2(n-m)} \ .
\end{align}

\end{proof}

\section{Remarks on integrals over unitary groups}
\label{NotesInt}
As previously mentioned, our results can be used to compute group integrals, i.e. integrals of the form $\int f(U,U^*) dU$ where one integrates over the entire group $\mathcal{U}(d)$ or $\mathcal{SU}(d)$, respectively. At least three things are needed when one intends to \emph{explicitly} compute such integrals: A parameterization of the corresponding group, exact knowledge of the parameter ranges and the normalized Haar measure. All this is provided in the present paper. Theorems \ref{theorem1} -- \ref{theorem4} can straightforwardly be applied without knowledge of further technicalities (e.g. details appearing in the proofs). Whether or not a given integral can be solved analytically in this way of course depends on the integrand. However, for many physical problems the function $f(U,U^*)$ is a polynomial in the components of $U$ and $U^{*}$. In this case, when $U$ and $dU$ are inserted in parameterized form according to Theorems \ref{theorem1} -- \ref{theorem4}, we have that the integrand is a polynomial in $\cos \lambda_{m,n}$, $\sin \lambda_{m,n}$ and $e^{\pm i\lambda_{m,n}}$. Neglecting the computational effort, such integrals can always be solved analytically (see Ref.~\cite{Gradst} and references therein). Besides that, our results constitute a good starting point for the integration of non-polynomial functions using numerical methods.

A detailed analysis on integrals that can be solved in this way shall be presented in a subsequent paper. We are convinced that due to the simplicity of the parameterization and the associated Haar measure, it is possible to find several general results and to gain a better understanding of integrals over $\mathcal{U}(d) \ / \ \mathcal{SU}(d)$. In order not to go beyond the scope of this paper we conclude with simple examples that can be compared with existing results as a consistency check.
\subsection{Example 1}
In Ref.~\cite{Collins} it was shown that
\begin{align}
\label{collinsrel}
\int_{\mathcal{U}(d)} |\bra{1}U\ket{1}|^4 dU=\frac{2}{d(d+1)} \ .
\end{align}
By means of our results, it is straightforward to analytically confirm this relation and to find the general solution for $\int |\bra{1}U\ket{1}|^p dU$ for arbitrary $p \in \mathbb{N}$. In parameterized form we have $\bra{1}U_C\ket{1}=e^{i \lambda_{1,1}} \prod_{n=2}^{d} \cos(\lambda_{1,n})$, and accordingly $|\bra{1}U_C\ket{1}|^p=\prod_{n=2}^{d} \cos^p(\lambda_{1,n})$ only depends on $\lambda_{1,2},\ldots,\lambda_{1,d}$. Due to this, the integral simplifies as follows (see also Appendix \ref{diffmatrix})
\begin{align}
\label{bintegral}
&\int_{\mathcal{U}(d)} |\bra{1}U\ket{1}|^p dU \ , \\
=&\int_{\lambda_{k,l}=0}^{2\pi \ (k\geq l)}\int_{\lambda_{k,l}=0}^{\pi/2 \ (k<l)} \prod_{n=2}^{d} \cos^p(\lambda_{1,n}) \ dU_d \ ,  \\
=&\int_{\lambda_{1,2}=0}^{\pi/2}\cdots\int_{\lambda_{1,d}=0}^{\pi/2} \prod_{n=2}^{d} \cos^p(\lambda_{1,n}) 2(n-1) \sin(\lambda_{1,n})\cos^{2(n-1)-1}(\lambda_{1,n}) d\lambda_{1,2} \cdots d\lambda_{1,d} \ , \nonumber \\
=&\int_{\lambda_{1,2}=0}^{\pi/2}\cdots\int_{\lambda_{1,d}=0}^{\pi/2} \prod_{n=2}^{d} 2(n-1) \sin(\lambda_{1,n})\cos^{2(n-1)-1+p}(\lambda_{1,n}) d\lambda_{1,2} \cdots d\lambda_{1,d}  \ .
\end{align}
This integral can easily be solved, i.e.
 \begin{align}
 =&\left. \left[\prod_{n=2}^{d} 2(n-1) \frac{-\cos^{2(n-1)+p}(\lambda_{1,n})}{2(n-1)+p}\right] \right|_{\lambda_{1,n}=0}^{\pi/2}  \ , \\
 =&\prod_{n=2}^{d}  \frac{2(n-1)}{2(n-1)+p} \ .
 \label{eintegral}
\end{align}
For the special case $p=4$ the general solution
\begin{align}
\int_{\mathcal{U}(d)} |\bra{1}U\ket{1}|^p dU=\prod_{n=2}^{d}  \frac{2(n-1)}{2(n-1)+p}
\end{align}
simplifies to $\prod_{n=2}^{d}  \frac{2(n-1)}{2(n+1)}=\frac{2}{d(d+1)}$ as in agreement with Ref.~\cite{Collins}. Note that in this way we have found a simple necessary criterion for testing if a set of matrices constitutes a unitary design. Namely, as there are no distinguished matrix elements and since $|\bra{1}U\ket{1}|^{2t}=\bra{1}U\ket{1}^{t}\bra{1}U^*\ket{1}^{t}$ it holds: A set of unitaries $\{ U_i \}_N$ is a unitary $t$-design only if
\begin{align}
\sum_{i=1}^{N} w_i |\bra{k} U_i \ket{l}|^{2t} =\prod_{n=2}^{d}  \frac{n-1}{n-1+t}
\end{align}
for all $k, l \in \{1,\ldots,d\}$, where $w_i$ is the weighting of $U_i$ ($w_i=\frac{1}{N}$ for unweighted designs). Further criteria can be constructed analogously.
\subsection{Example 2}
It is known that bilateral twirling \mbox{$\int U \otimes U \rho U^{\dagger} \otimes U^{\dagger} dU$} of a state $\rho$ on $\mathbb{C}^d \otimes \mathbb{C}^d$ results in a \emph{Werner state} \cite{Werner,Eggeling,Chruscinski,Clarisse} which has the form
\begin{align}
\rho_W=\frac{\mathbbm{1} + \beta (\sum_{i,j=1}^{d} \ket{ij}\bra{ji})}{d(d+\beta)} \ ,
\end{align}
where $-1 \leq \beta \leq 1$. Using a symbolic computation software, we explicitly and analytically twirled the maximally entangled state $\ket{\Psi}=\frac{1}{\sqrt{d}} \sum_{i=1}^{d} \ket{i}\otimes \ket{i}$ of dimensions $d=2,3,4,5$ utilizing Theorem \ref{theorem1} and \ref{theorem3}
\begin{align}
\int_{\lambda_{k,l}=0}^{2\pi \ (k\geq l)}\int_{\lambda_{k,l}=0}^{\pi/2 \ (k<l)}  U_C \otimes U_C \ket{\Psi}\bra{\Psi} U_C^{\dagger} \otimes U_C^{\dagger} dU_d \ ,
\end{align}
and alternatively with Theorem \ref{theorem2} and \ref{theorem4}
\begin{align}
\int_{\lambda_{k,l}=0}^{2\pi \ (k=l)}\int_{\lambda_{k,l}=0}^{\pi \ (k> l)}\int_{\lambda_{k,l}=0}^{\pi/2 \ (k<l)} U_C \otimes U_C \ket{\Psi}\bra{\Psi} U_C^{\dagger} \otimes U_C^{\dagger} dU_d \ .
\end{align}
In all cases, this yielded a \emph{Werner state} with $\beta=1$, demonstrating once more the operationality and validity of our results. Note that our results also enable analytical twirling of multipartite qudit states.
\subsection{Example 3}
\begin{figure}[h]
\centering
\setlength{\unitlength}{1cm}
  \begin{picture}(10,6.5)(0,0)
  \put(0,0){\includegraphics[scale=1]{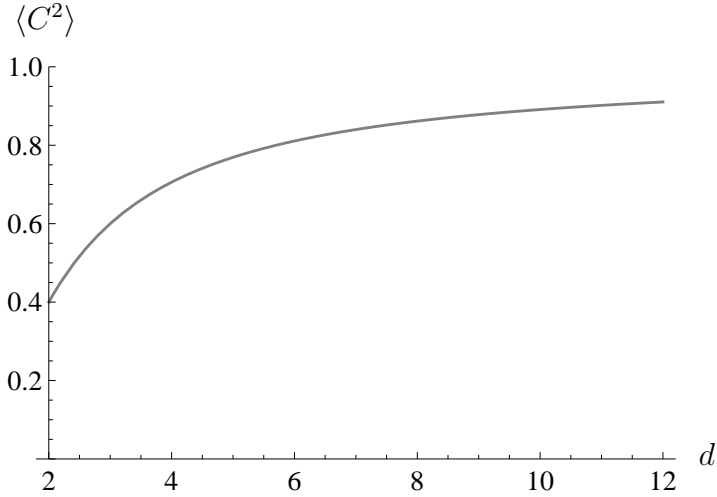}}
  \put(0.5,6.3){\fontsize{12}{12}\selectfont \makebox(0,0)[]{$\langle C^2 \rangle$\strut}}
  \put(9.2,0.5){\fontsize{12}{12}\selectfont \makebox(0,0)[]{$d$\strut}}
   \end{picture}
  \caption{The average entanglement (\ref{apriorient}) of a bipartite qudit system measured by the (squared) concurrence $C^2(\ket{\psi})$ for the dimensions $d=2,\ldots,12$. The average $\langle C^2 \rangle$ increases with the size of the Hilbert space $\mathcal{H}_A \otimes \mathcal{H}_B=\mathbb{C}^d \otimes \mathbb{C}^d$.}
  \label{averagecon}
\end{figure}

The entanglement of a bipartite qudit state $\ket{\psi} \in \mathcal{H}_A \otimes \mathcal{H}_B = \mathbb{C}^d \otimes \mathbb{C}^d$ can be quantified via the (normalized) concurrence \cite{conc1,conc2,conc3,conc4} determined by $C^2(\ket{\psi})=\frac{d}{d-1}(1-\Tr(\rho_B^2))$, where $\rho_B$ is the reduced density matrix $\rho_B=\Tr_A(\ket{\psi}\bra{\psi})$. An interesting property of a quantum system is the \emph{a priori entanglement}, i.e. a characteristic such as the generic probability that a state is entangled or the average amount of entanglement over all states in dependence of the dimension $d$. Here, we focus on the average
\begin{align}
\langle C^2 \rangle=\int_{\mathcal{U}(d^2)}C^2(U \ket{\psi}) dU \ ,
\end{align}
where $\ket{\psi}$ is an arbitrary state of $\mathbb{C}^d \otimes \mathbb{C}^d$. If we express $\ket{\psi'}\bra{\psi'}=U \ket{\psi}\bra{\psi} U^{\dagger}$ as $\ket{\psi'}\bra{\psi'}=\sum_{i,j=1}^{d^2} u_{i}u^{*}_{j} \ket{i}\bra{j}$ we obtain the reduced density matrix
\begin{align}
\label{reddensity}
\rho_B=\sum_{i,j=1}^{d} \left(\sum_{n=0}^{d-1} u_{i+nd}u^{*}_{j+nd} \right)\ket{i}\bra{j} \ .
\end{align}
To solve the integral $\int \frac{d}{d-1}(1-\Tr(\rho_B^2))dU$ we can associate each $u_{i}$ with a matrix element of a $d^2 \times d^2$ unitary matrix. In this way, the average $\langle C^2 \rangle$ reduces to a sum of integrals over the polynomials $I_4=\int |\bra{k}U\ket{l}|^4 dU$ and $I_{2,2}=\int |\bra{k}U\ket{l}|^2|\bra{m}U\ket{n}|^2 dU$ where $\bra{k}U\ket{l}$ and $\bra{m}U\ket{n}$ denote distinct matrix elements. Now, taking account of (\ref{reddensity}), it is a simple combinatorial problem to show that in $\int \Tr(\rho_B^2)dU$ the term $I_4$ appears $d^2$ times and $I_{2,2}$ appears $2(d-1)d^2$ times. Hence,
\begin{align}
\langle C^2 \rangle=\frac{d}{d-1}(1-d^2I_4-2(d-1)d^2I_{2,2}) \ .
\end{align}
From (\ref{collinsrel}) we already know that $I_4=\frac{2}{d^2(d^2+1)}$ for integrals over $\mathcal{U}(d^2)$. It remains to determine $I_{2,2}$ for two arbitrary but distinct matrix elements $\bra{k}U\ket{l}$ and $\bra{m}U\ket{n}$. By computing
\begin{align}
I_{2,2}=\int_{\mathcal{U}(d^2)} |\bra{1}U_C\ket{1}|^2|\bra{d^2}U_C\ket{1}|^2 dU_{d^2} \ ,
\end{align}
analogously to (\ref{bintegral}) -- (\ref{eintegral}) one finds with $|\bra{d^2} U_C \ket{1}|^2=\sin^2(\lambda_{1,d^2})$ that $I_{2,2}=\frac{1}{d^2(d^2+1)}=\frac{1}{2}I_4$. Consequently, the average entanglement of a bipartite qudit system is
\begin{align}
\langle C^2 \rangle=\frac{d(d-1)}{d^2+1} \ .
\label{apriorient}
\end{align}
This result is graphically depicted in Figure~\ref{averagecon}. Physically interpreted it means that the higher-dimensional the system is, the more likely it becomes to obtain a highly entangled state when picking a pure state of $\mathcal{H}=\mathbb{C}^d \otimes \mathbb{C}^d$ at random.

\section{Summary}
In this paper we adopted the concept of the composite parameterization of the unitary group $\mathcal{U}(d)$ to the special unitary group $\mathcal{SU}(d)$. We showed that both parameterizations can be used equivalently to describe orthonormal vectors and subspaces with the minimal number of parameters. The introduced parameterizations are completely factorized and therefore beneficial for numerical optimizations. We also determined the infinitesimal volume element in terms of the introduced parameters. We derived a general formula of the normalized Haar measure for both the unitary $\mathcal{U}(d)$ and the special unitary group $\mathcal{SU}(d)$ of arbitrary dimension. The found expressions give theoretical insights into the differential structure of $\mathcal{U}(d)$ and $\mathcal{SU}(d)$. Moreover, the Haar measure plays an important role in all kinds of unbiased randomizations. It was stressed that our results also constitute a framework for computing high-order group integrals. By means of our approach, we analytically solved several exemplary integrals and found that the solutions are in agreement with the literature. As integrals over unitary groups appear in various fields from particle physics to quantum optics to quantum information, it is to be expected that our results will find several interesting applications.
\vspace{0.5cm}\\
\textbf{Acknowledgments:}
The authors want to thank Jan Bouda, Artem Kaznatcheev, Patrick Ludl, David Rottensteiner and the Anonymous Referee for helpful remarks and their valuable comments on the manuscript. Christoph Spengler and Marcus Huber acknowledge financial support from the Austrian Science Fund (FWF) - Project P21947N16.
\appendix
\section{Some explicit expressions for $\mathcal{U}(d)$}
$d=2$ :
\begin{align*}
U_C \ = \ &\exp \left( i P_2 \lambda_{2,1} \right) \exp \left( i Y_{1,2} \lambda_{1,2} \right) \exp \left( i P_1 \lambda_{1,1} \right) \exp \left( i P_2 \lambda_{2,2} \right) \ , \\
dU_2 \ = \ &\frac{1}{4 \pi ^3} \sin \lambda_{1,2} \cos \lambda_{1,2} \ d\lambda_{1,2} d\lambda_{2,1} d\lambda_{1,1} d\lambda_{2,2}
\end{align*}
$d=3$ :
\begin{align*}
U_C \ = \ &\exp \left( i P_2 \lambda_{2,1} \right) \exp \left( i Y_{1,2} \lambda_{1,2} \right)\exp \left( i P_3 \lambda_{3,1} \right) \exp \left( i Y_{1,3} \lambda_{1,3} \right) \exp \left( i P_3 \lambda_{3,2} \right) \\
  & \times \exp \left( i Y_{2,3} \lambda_{2,3} \right) \exp \left( i P_1 \lambda_{1,1} \right) \exp \left( i P_2 \lambda_{2,2} \right)\exp \left( i P_3 \lambda_{3,3} \right) \ , \\
dU_3 \ = \ &\frac{1}{4 \pi ^6} \sin \lambda_{1,2} \cos \lambda_{1,2} \sin \lambda_{1,3} \cos^3 \lambda_{1,3} \sin \lambda_{2,3} \cos \lambda_{2,3} \\
& \times d\lambda_{1,2} d\lambda_{2,1} d\lambda_{1,3} d\lambda_{3,1} d\lambda_{2,3} d\lambda_{3,2} d\lambda_{1,1} d\lambda_{2,2} d\lambda_{3,3}
\end{align*}
$d=4$ :
\begin{align*}
U_C \ = \ &\exp \left( i P_2 \lambda_{2,1} \right) \exp \left( i Y_{1,2} \lambda_{1,2} \right)\exp \left( i P_3 \lambda_{3,1} \right) \exp \left( i Y_{1,3} \lambda_{1,3} \right) \exp \left( i P_4 \lambda_{4,1} \right)  \\
& \times \exp \left( i Y_{1,4} \lambda_{1,4} \right) \exp \left( i P_3 \lambda_{3,2} \right) \exp \left( i Y_{2,3} \lambda_{2,3} \right)\exp \left( i P_4 \lambda_{4,2} \right) \exp \left( i Y_{2,4} \lambda_{2,4} \right) \\
& \times \exp \left( i P_4 \lambda_{4,3} \right) \exp \left( i Y_{3,4} \lambda_{3,4} \right) \exp \left( i P_1 \lambda_{1,1} \right) \exp \left( i P_2 \lambda_{2,2} \right)\exp \left( i P_3 \lambda_{3,3} \right) \exp \left( i P_4 \lambda_{4,4} \right) \ , \\
dU_4 \ = \ &\frac{3}{4 \pi ^{10}}  \sin \lambda_{1,2} \cos \lambda_{1,2} \sin \lambda_{1,3} \cos^3 \lambda_{1,3} \sin \lambda_{1,4} \cos^5 \lambda_{1,4} \sin \lambda_{2,3} \cos \lambda_{2,3} \\
& \times \sin \lambda_{2,4} \cos^3 \lambda_{2,4} \sin \lambda_{3,4} \cos \lambda_{3,4} \ d\lambda_{1,2} d\lambda_{2,1} d\lambda_{1,3} d\lambda_{3,1} d\lambda_{1,4} d\lambda_{4,1} \\
& \times d\lambda_{2,3} d\lambda_{3,2} d\lambda_{2,4} d\lambda_{4,2} d\lambda_{3,4} d\lambda_{4,3} \lambda_{1,1} d\lambda_{2,2} d\lambda_{3,3} d\lambda_{4,4}
\end{align*}
\section{Some explicit expressions for $\mathcal{SU}(d)$}
$d=2$ :
\begin{align*}
U_C \ = \ &\exp \left( i Z_{1,2} \lambda_{2,1} \right) \exp \left( i Y_{1,2} \lambda_{1,2} \right) \exp \left( i Z_{1,2} \lambda_{1,1} \right) \ , \\
dU_2 \ = \ &\frac{1}{\pi ^2} \sin \lambda_{1,2} \cos \lambda_{1,2} \ d\lambda_{1,2} d\lambda_{2,1} d\lambda_{1,1}
\end{align*}
$d=3$ :
\begin{align*}
U_C \ = \ &\exp \left( i Z_{1,2} \lambda_{2,1} \right) \exp \left( i Y_{1,2} \lambda_{1,2} \right)\exp \left( i Z_{1,3} \lambda_{3,1} \right) \exp \left( i Y_{1,3} \lambda_{1,3} \right) \\
& \times \exp \left( i Z_{2,3} \lambda_{3,2} \right) \exp \left( i Y_{2,3} \lambda_{2,3} \right) \exp \left( i Z_{1,3} \lambda_{1,1} \right) \exp \left( i Z_{2,3} \lambda_{2,2} \right) \ , \\
dU_3 \ = \ &\frac{4}{\pi ^5} \sin \lambda_{1,2} \cos \lambda_{1,2} \sin \lambda_{1,3} \cos^3 \lambda_{1,3} \sin \lambda_{2,3} \cos \lambda_{2,3}\\
& \times  d\lambda_{1,2} d\lambda_{2,1} d\lambda_{1,3} d\lambda_{3,1} d\lambda_{2,3} d\lambda_{3,2} d\lambda_{1,1} d\lambda_{2,2}
\end{align*}
$d=4$ :
\begin{align*}
U_C \ = \ &\exp \left( i Z_{1,2} \lambda_{2,1} \right) \exp \left( i Y_{1,2} \lambda_{1,2} \right)\exp \left( i Z_{1,3} \lambda_{3,1} \right) \exp \left( i Y_{1,3} \lambda_{1,3} \right) \exp \left( i Z_{1,4} \lambda_{4,1} \right) \\
& \times \exp \left( i Y_{1,4} \lambda_{1,4} \right) \exp \left( i Z_{2,3} \lambda_{3,2} \right) \exp \left( i Y_{2,3} \lambda_{2,3} \right)\exp \left( i Z_{2,4} \lambda_{4,2} \right) \exp \left( i Y_{2,4} \lambda_{2,4} \right)\\
& \times \exp \left( i Z_{3,4} \lambda_{4,3} \right) \exp \left( i Y_{3,4} \lambda_{3,4} \right) \exp \left( i Z_{1,4} \lambda_{1,1} \right) \exp \left( i Z_{2,4} \lambda_{2,2} \right)\exp \left( i Z_{3,4} \lambda_{3,3} \right)\\
dU_4 \ = \ &\frac{96}{\pi ^9}  \sin \lambda_{1,2} \cos \lambda_{1,2} \sin \lambda_{1,3} \cos^3 \lambda_{1,3} \sin \lambda_{1,4} \cos^5 \lambda_{1,4} \sin \lambda_{2,3} \cos \lambda_{2,3}\\
& \times  \sin \lambda_{2,4} \cos^3 \lambda_{2,4} \sin \lambda_{3,4} \cos \lambda_{3,4} \ d\lambda_{1,2} d\lambda_{2,1} d\lambda_{1,3} d\lambda_{3,1} d\lambda_{1,4} d\lambda_{4,1}\\
& \times  d\lambda_{2,3} d\lambda_{3,2} d\lambda_{2,4} d\lambda_{4,2} d\lambda_{3,4} d\lambda_{4,3} \lambda_{1,1} d\lambda_{2,2} d\lambda_{3,3}
\end{align*}
\section{Differential matrix representation for $\mathcal{U}(d)$}
\label{diffmatrix}
Adopting the matrix representation $[ \lambda_{m,n} ]$ introduced in Section \ref{remarksUSU} to differentials, the normalized Haar measure may be written as
\begin{align}
\label{harald}
dU_d=\prod_{m,n=1}^{d} \Delta_{m,n} \ ,
\end{align}
with
\begin{align}
\label{haardelta}
 \left[ \Delta_{m,n} \right]=\left[
  \begin{array}{ccccc}
    \frac{d\lambda_{1,1}}{2\pi}  & -d ( \cos^2 (\lambda_{1,2}) ) & \cdots & -d ( \cos^{2(d-1)} (\lambda_{1,d}) ) \\
    \frac{d\lambda_{2,1}}{2\pi} & \ddots & \ddots & \vdots \\
    \vdots & \ddots& \ddots &  -d ( \cos^2 (\lambda_{d-1,d}) )  \\
         \frac{d\lambda_{d,1}}{2\pi}  & \cdots & \frac{d\lambda_{d,d-1}}{2\pi}    & \frac{d\lambda_{d,d}}{2\pi}   \\
   \end{array}
\right] \ .
\end{align}
This notation is useful for the construction of a normalized Haar measure for problems which are independent of certain parameters $\lambda_{k,l}$ (see discussion in Section \ref{remarksUSU} and Ref.~\cite{Composite}, as well as \emph{Example 1}). Since in such cases the dependence on $\lambda_{k,l}$ and $d \lambda_{k,l}$ can be removed from the Haar measure, one can replace the corresponding entry $\Delta_{k,l}$ by a constant. If one sets $\Delta_{k,l}=1$ then (\ref{harald}) preserves the normalization of the reduced Haar measure. Moreover, this notation could lead to a better understanding of the differential geometry of $\mathcal{U}(d)$.
\section{Differential matrix representation for $\mathcal{SU}(d)$}
Analogously, the normalized Haar measure on $\mathcal{SU}(d)$ may be written as
\begin{align}
\label{haaralt}
dU_d=\prod_{m,n=1}^{d} \Delta_{m,n} \ ,
\end{align}
with
\begin{align}
 \left[ \Delta_{m,n} \right]=\left[
  \begin{array}{ccccc}
    \frac{d\lambda_{1,1}}{2\pi}  & -d ( \cos^2 (\lambda_{1,2}) ) & \cdots & -d ( \cos^{2(d-1)} (\lambda_{1,d}) ) \\
    \frac{d\lambda_{2,1}}{\pi} & \ddots & \ddots & \vdots \\
    \vdots & \ddots& \frac{d\lambda_{d-1,d-1}}{2\pi}  &  -d ( \cos^2 (\lambda_{d-1,d}) )  \\
         \frac{d\lambda_{d,1}}{\pi}  & \cdots & \frac{d\lambda_{d,d-1}}{\pi}    &  1 \\
   \end{array}
\right] \ .
\end{align}

\end{document}